\documentclass[11pt]{article}

\usepackage[margin=1in]{geometry}
\usepackage{fixltx2e,fix-cm}
\usepackage{amssymb}
\usepackage{amsmath}
\usepackage{graphicx}
\usepackage{subfigure}
\usepackage{makeidx}
\usepackage{multicol}

\usepackage{epsfig}
\usepackage{hyperref} 
\usepackage{setspace}

\makeindex

\usepackage{natbib}

\usepackage[sectionbib]{chapterbib}

\usepackage{booktabs}

\newcommand{\be}{{\boldsymbol e}}

\newcommand{\bt}{{\boldsymbol t}}
\newcommand{\bm}{{\boldsymbol m}}
\newcommand{\bu}{{\boldsymbol u}}
\newcommand{\bv}{{\boldsymbol v}}
\newcommand{\bw}{{\boldsymbol w}}
\newcommand{\bx}{{\boldsymbol x}}
\newcommand{\bz}{{\boldsymbol z}}
\newcommand{\by}{{\boldsymbol y}}
\newcommand{\bX}{{\boldsymbol X}}
\newcommand{\bY}{{\boldsymbol Y}}
\newcommand{\bM}{{\boldsymbol M}}
\newcommand{\bP}{{\boldsymbol P}}

\newcommand{\bW}{{\boldsymbol W}}
\newcommand{\ba}{{\boldsymbol a}}
\newcommand{\bb}{{\boldsymbol b}}
\newcommand{\btheta}{{\boldsymbol \theta}}
\newcommand{\bmu}{{\boldsymbol \mu}}

\newcommand{\binf}{{\boldsymbol \infty}}
\newcommand{\bzero}{{\boldsymbol 0}}
\newcommand{\der}{{\mathrm d}}
\def\cA{\mathcal{A}}
\def\cB{\mathcal{B}}
\def\sA{\mathbb{A}}
\def\cS{\mathcal{S}}
\def\sW{\mathbb{W}}
\def\cW{\mathcal{W}}
\def\sS{\mathbb{S}}

\def\cF{\mathcal{F}}
\def\sP{\mathbb{P}}
\def\cL{\mathcal{L}}
\def\P{\textrm{pr}}
\def\E{\textrm{E}}
\newcommand{\real}{{\mathbb R}}
\newcommand{\nat}{{\mathbb N}}
\def\indic{\textrm{1\!I}}

\DeclareMathOperator{\tr}{tr}
\DeclareMathOperator{\var}{Var}

\begin{document}
\title{Extreme Dependence Models}

\author{Boris B\'{e}ranger
\thanks{Theoretical and Applied Statistics Laboratory (LSTA),
University Pierre and Marie Curie - Paris 6, F-75005, Paris, France}
\thanks{School of Mathematics and Statistics, 
University of New South Wales, Sydney, Australia}, 
Simone A. Padoan
\thanks{Department of Decision Sciences, Bocconi University of Milan, 20136 Milano, Italy}
}

\date{}

\maketitle
\section*{Abstract}

Extreme values of real phenomena are events that occur with low frequency, 
but can have a large impact on real life.
These are, in many practical problems, high-dimensional by nature 
\citep[e.g.][]{tawn1990, coles1991}.
To study these events is of fundamental importance. For this purpose, probabilistic models and statistical methods are in high demand. 
There are several approaches to modelling multivariate extremes as described in \citet{falk2011},
linked to some extent. 
We describe an approach for deriving multivariate extreme value models and we illustrate the main features
of some flexible extremal dependence models. 
We compare them by showing their utility with a real data application, in particular 
analyzing the extremal dependence among several pollutants recorded in the city of Leeds, UK.
%

\section{Introduction}\label{sec:intro}

Statistical analyses of extreme events are of crucial importance for risk assessment in many areas such as 
the financial market, telecommunications, industry, environment and health. For example governments and insurance companies need to statistically quantify the frequency of natural disasters in order to plan risk management and take preventive actions. 

Several examples of univariate analysis are available, for instance in \citet{coles2001}.
Two main approaches are used in applications, the block-maximum and the peak over a threshold. These are based on the generalized extreme value (GEV) distribution and the generalized Pareto distribution (GPD), which are milestones of the extreme value theory, see e.g. \citet[][Ch. 3--4]{coles2001} and the references therein.

Many practical problems in finance, the environment, etc. are high-dimensional by nature, for example when analyzing the air quality in an area, the amount of pollution depends on the levels of different pollutants and the interaction between them. 
Today the extreme value theory provides a sufficiently mature framework for applications in the multivariate case. Indeed a large number of theoretical results and statistical methods and models are available, see for instance the monographs \citet{resnick2007}, \citet{dehaan2006}, \citet{falk2011}, \citet{beirlant2006}, \citet{coles2001} and \citet{kotz2000}. 
In this article we review some basic theoretical results on the extreme values of multivariate variables 
(multivariate extremes for brevity). 
With the block-maximum approach we explain what type of dependence structures can be described.
We discuss the main features of some families of parametric extremal dependence models. 
By means of real data analysis we show the utility of these extremal dependence models when assessing the dependence of multivariate extremes. Their utility is also illustrated when estimating the probabilities that multivariate extreme events occur.

The analysis of real phenomena such as heavy rainfall, heat waves and so on is a challenging task. 
The first difficulty is the complexity of the data, i.e. observations are collected over space and time. 
In this case, theory deals with extremes of temporal- or spatial-processes \citep[e.g.][Ch. 9]{dehaan2006}. 
Examples of such statistical analysis are \citet{davison2012b},
\cite{davison2012c}, for a simple review see \citet{padoan2013a}.
This theory is closely linked to that of multivariate extremes presented here. 
The second difficulty is that the dependence
of multivariate extremes is not always well captured by the models illustrated here.
\citet{ledford+t1996, ledford1997} have shown that in some applications a more suitable dependence structure is described by the so called {\it asymptotic independence}.
This framework has been recently extended to continuous processes \citep[e.g.][]{de2011, wadsworth+t12, padoan2013c}.
These motivations make the multivariate extreme value theory a very active research field at present.

The paper is organized as follows. In Section 1.2 a definition of multivariate extremes is provided and the
main characteristics are presented. In Section 1.3 some of the most popular extremal dependence models
are described. In Section 1.4 some estimation methods are discussed and in Section 1.5 the analysis of
the extremes of multiple pollutants is performed.

\section{Multivariate Extremes}

Applying the block-maximum approach to every component of a multivariate random vector gives rise to a definition of multivariate extremes.
Specifically, for $d\in\nat$, let $I=\{1,\ldots,d\}$ be an index set and 
$\bX=( X_{1}, \ldots, X_{d} )$ be an $\real^d$-valued random vector with joint (probability) distribution function $F$ and marginal distribution functions
$F_j=F(\infty,\ldots,x_j,\ldots,\infty)$, $j \in I$.
\index{block-maxima}
Suppose that $\bX_1,\ldots,\bX_n$ are $n$ independent and identically distributed (i.i.d.) copies of $\bX$.
The sample vector of componentwise maxima (sample maxima for brevity) is 
$\bM_n = (M_{n,1}, \ldots, M_{n,d})$, where $M_{n,j}=\max(X_{1,j},\ldots,X_{n,j})$.
\index{Component wise maxima}

Typically, in applications the distribution $F$ is unknown and so the distribution of 
the sample maxima is also unknown. 
A possible solution is to study the asymptotic distribution of $M_n$ as $n\rightarrow \infty$ and to use it as an approximation for a large but finite sample size, resulting in an approximate distribution for multivariate extremes.
At a first glance, this notion of multivariate extremes may seem too simple to provide a useful approach for applications. However, a number of theoretical results justify its practical use.
For example, with this definition of multivariate extremes, the dependence that arises is linked to the dependence that all the components of $\bX$ are simultaneously large. 
Thus, by estimating these dependence structures we are also able to estimate the probabilities that multiple exceedances occur.
\subsection{Multivariate extreme value distributions}\label{ss:maxstab}

The asymptotic distribution of $\bM_n$ is derived with a similar
approach to the univariate case.
Assume there are sequences of normalizing constants $\ba_n = (a_{n1}, \ldots, a_{nd})>\bzero$, with 
$\bzero=(0,\ldots,0)$, and $\bb_n = ( b_{n1}, \ldots, b_{nd} ) \in \real^d$ such that
\begin{align}\label{eq:limMEVD}
\P\bigg( \frac{\bM_n - \bb_n}{\ba_n} \leq \bx \bigg) 
= F^n(\ba_n \bx + \bb_n) \rightarrow G(\bx),
\quad\quad n\rightarrow \infty, 
\end{align} 
for all the continuity points $\bx$ of a non-degenerate distribution $G$. 
The class of the limiting distributions in \eqref{eq:limMEVD} is called
{\it multivariate extreme value distributions} (MEVDs) \citep[][p.~263]{resnick2007}.
A distribution function $F$ that satisfies the convergence result \eqref{eq:limMEVD} 
is said to be in the {\it (maximum) domain of attraction} 
of $G$ \citep[][pp.~226--229]{dehaan2006}.
An attractive property of MEVDs is the {\it max-stability}.
\index{Max-stable}
A distribution $G$ on $\real^d$ is max-stable if for every $n\in\nat$, there exists sequences $\ba_n>\bzero$ and
$\bb_n\in\real^d$ such that
\begin{equation}\label{eq:stability}
G(\ba_n\,\bx+\bb_n)=G^{1/n}(\bx),
\end{equation}
\citep[Proposition~5.9]{resnick2007}.
As a consequence, $G$ is such that $G^a$ is a distribution for every $a>0$.
A class of distributions that satisfies such a property is named {\it max-infinitely divisible} (max-id).
\index{Max-id} 
More precisely, a distribution $G$ on $\real^d$
is max-id, if for any $n\in\nat$ there exists a distribution $F_n$ such that $G=F^n_n$ 
\citep[][p.~252]{resnick2007}. 
This means that $G$ can always be defined
through the distribution of the sample maxima of $n$ i.i.d. random vectors.

In order to characterize the class of MEVDs we need to specify: a) the form of the marginal distributions,
b) the form of the dependence structure.

a) To illustrate the first feature is fairly straightforward. 
If $F$ converges, then so too does the marginal distributions $F_j$ for all $j \in I$.
Choosing $a_{jn}$ and $b_{jn}$ for all $j\in I$ as in \citet[][Corollary 1.2.4]{dehaan2006}, implies that each marginal distribution of $G$ is a generalized extreme value (GEV), i.e.
\index{Generalised extreme value distribution}
$$
G(\infty,\ldots,x_j,\ldots,\infty)=\exp
\left[
-\left\{
1+\xi_j
\left(
\frac{x_j-\mu_j}{\sigma_j}
\right)
\right\}_{+}^{-1/\xi_j}
\right],\; j\in I,
$$
where $(x)_+=\max(0,x)$, $-\infty<\mu_j,\xi_j<\infty$, $\sigma_j>0$ \citep[][pp. 208--211]{dehaan2006}. 
Because the marginal distributions are continuous then $G$ is also continuous. 

b) The explanation of the dependence form is more elaborate, although it is not complicated.
The explanation is based on three steps: 1) $G$ is transformed so that its marginal distributions are equal, 
2) a Poisson point process (PPP) is used to represent the standardised distribution,
3) the dependence form is made explicit by means of a change of coordinates.
Here are the steps.

1) Let $U_j(a)=F_j^{\leftarrow}(1-1/a)$, with $a>1$, be the left-continuous inverse of $F_j$, for all $j\in I$.
The sequences $a_{nj}$ and $b_{nj}$ in \eqref{eq:limMEVD} are such that
for all $y_j>0$,
$$
\lim_{n\rightarrow\infty} \frac{U_j(ny_j)-b_n}{a_n}=\frac{\sigma_j(y_j^{\xi_j}-1)}{\xi_j} + \mu_j,\quad j\in I,
$$
and therefore
\begin{align}\label{eq:limit_fre}
\lim_{n\rightarrow\infty}F^n \{ & U_1(ny_1), \ldots,U_d(ny_d)\} \nonumber\\
&=G\left(
\frac{\sigma_1(y_1^{\xi_1}-1)}{\xi_1}+\mu_1,\ldots,
\frac{\sigma_d(y_d^{\xi_d}-1)}{\xi_d}+\mu_d
\right)\equiv G_0(\by),
\end{align}
for all continuity points $\by>\bzero$ of $G_0$ \citep[see][Theorems~1.1.6, 6.1.1]{dehaan2006}. 
$G_0$ is a MEVD with identical unit Fr\'{e}chet marginal distributions.

Now, for all $\by>\bzero$ such
that $0<G_0(\by)<1$, by taking the logarithm on the right and left side of \eqref{eq:limit_fre} and
using a first order Taylor expansion of $\log F\{U_1(ny_1),\ldots,U_d(ny_d)\}$, as $n\rightarrow\infty$,
it follows that
\begin{equation}\label{eq:exp_fun}
\lim_{n\rightarrow\infty} n [1- F\{U_1(ny_1),\ldots,U_d(ny_d)\}]=
-\log G_0(\by)\equiv V(\by).
\end{equation}
The function $V$, named {\it exponent (dependence) function}, represents the
dependence structure of multiple extremes (extremal dependence for brevity).
\index{Exponent dependence function}
According to \eqref{eq:exp_fun} the derivation of $V$ depends on the functional form of $F$.
In most of the practical problems the latter is unknown.
A possible solution is obtained exploiting the max-id property of $G_0$, which says
that every max-id distribution permits a PPP representation, 
see \citet[][pp.~257--262]{resnick2007} and \citet[][pp.~141--142]{falk2011}.

2) Let $N_n(\cdot)$ be a PPP defined by
$$
N_n(\cA):=\sum_{i=1}^\infty \indic_{\{\bP_i\}}(\cA),\quad
\indic_{\{\bP_i\}}(\cA)=\left\{
\begin{array}[c]{cc}%
1, & \bP_i \in \cA,\\
0, & \bP_i \notin \cA,\\
\end{array}
\right.
$$
where $\cA\subset \sA$ with $\sA:=(0,\infty)\times \real_+^d$,
\begin{equation*}\label{eq:ppp}
\bP_i=\left[
\frac{i}{n}, \left\{1+\xi_1\left(\frac{X_{i1}-b_{n1}}{a_{n1}}\right)\right\}^{\frac{1}{\xi_1}},\ldots,
\left\{1+\xi_1\left(\frac{X_{id}-b_{nd}}{a_{nd}}\right)\right\}^{\frac{1}{\xi_d}}
\right], 
\end{equation*}
for every $n\in \nat$ and $\bX_i$, $i=1,2,\ldots$ are i.i.d random vectors with distribution $F$. 
The intensity measure is $\zeta\times \eta_n$ 
where $\zeta$ is the Lebesgue measure and for every $n\in \nat$ and all critical regions 
defined by $\cB_{\by}:=\real_+^d\backslash[\bzero,\by]$ with $\by>0$,
$$
\eta_n(\cB_{\by})=n [1- F\{U_1(ny_1),\ldots,U_d(ny_d)\}],
$$
is a finite measure.
If the limit in \eqref{eq:limit_fre} holds, 
then $N_n$ converges weakly to $N$ as $n\rightarrow\infty$, i.e. a PPP with intensity measure $\zeta\times \eta$ where
$$
\eta(\cB_{\by})=\eta\{(\bv\in\real^d_+:v_1>y\,\text{or}\ldots\text{or}\,v_d>y_d)\}\equiv V(\by),\quad \by>0,
$$
is a fine measure, named {\it exponent measure} \citep[see][Theorems~6.1.5, 6.1.11]{dehaan2006}.
\index{Exponent measure}
Observe that $\eta$ must concentrate on $\overline{\real}=\real_+^d\backslash\{\bzero\}$
in order to be uniquely determined. Also, $\eta$ must satisfy $\eta(\binf)=0$, see
\citet[][p.~143]{falk2011} for details.

\begin{figure}[t!]
\begin{center}$
\begin{array}{c}
\includegraphics[width=0.8\textwidth]{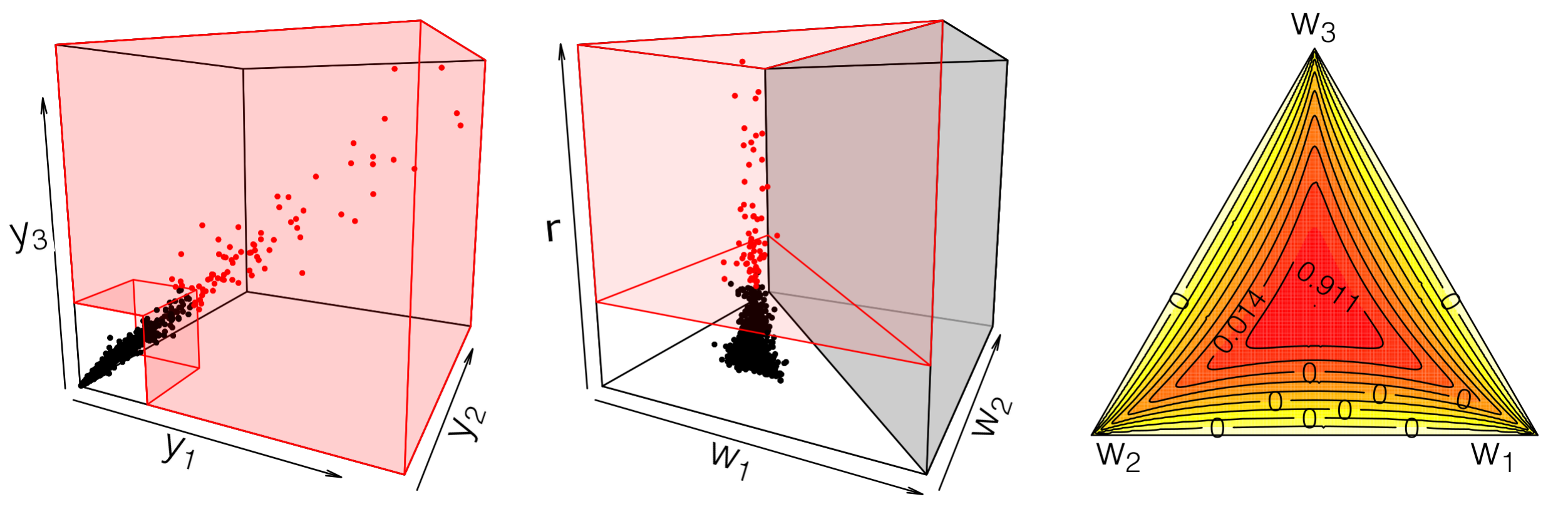}\\
\includegraphics[width=0.8\textwidth]{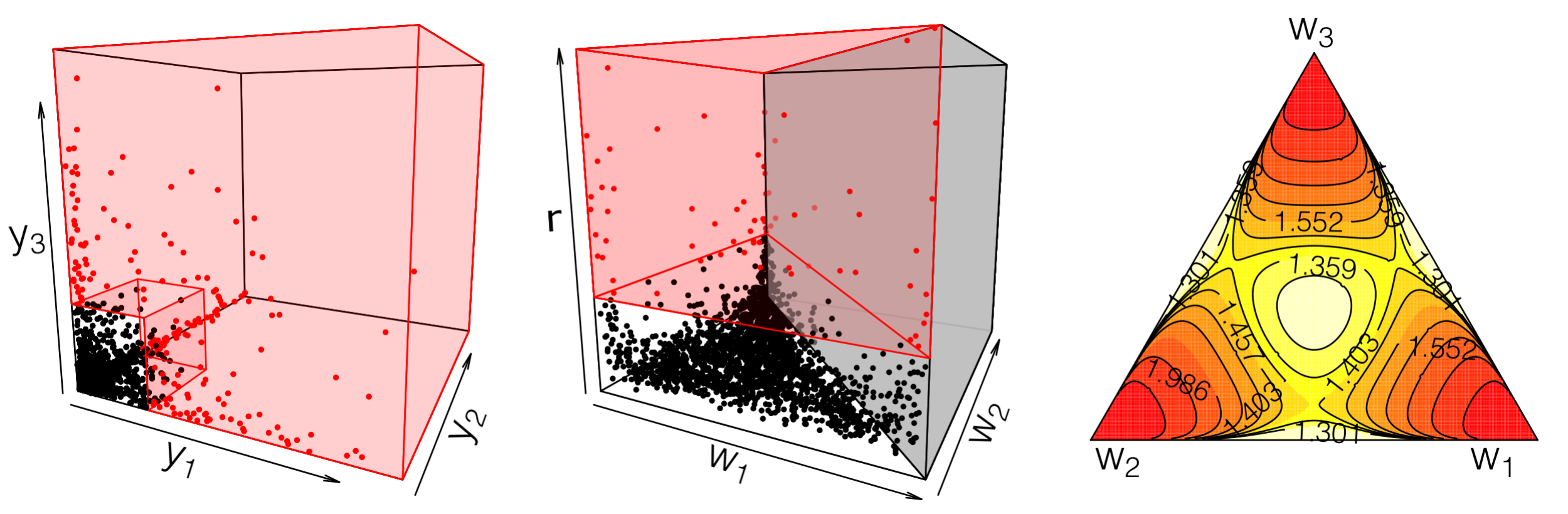}
\end{array}$
\end{center}
\caption
[Examples of critical regions and angular densities]
{Examples of critical regions in $\real_+^3$ (left-panels) and its representation in
pseudo-polar coordinates (middle-panels). Red points are the extremes with strong (top-panels)
and weak (bottom-panels) dependence. Right panels display the angular densities on the simplex.}
\label{fig:extremal_dep}
\end{figure}

This essentially means that numbering the rescaled observations that fall in a critical region, 
e.g. see the shaded sets in the left panels of Figure \ref{fig:extremal_dep}, 
where at least one coordinate is large, makes it possible for \eqref{eq:limit_fre} to be computed using the void probability of 
$N$, that is 
\begin{equation}\label{eq:prob_maxstab}
\begin{split}
G_0(\by)&=\P[N\{(0,1]\times \cB_{\by}\}=0]\\
&=\exp(-[\zeta\{(0,1]\}\times \eta(\cB_{\by})])\\
&=\exp\{-V(\by)\}\quad \by>\bzero.
\end{split}
\end{equation}
From Figure \ref{fig:extremal_dep} we see that in the case of strong dependence (top-left panel) all the coordinates of the extremes are large, while in the case of weak dependence (bottom-left panels) only one coordinate of the extremes is large. 

At this time it remains to be specify the structure of the exponent measure. This task is simpler to fulfil when working with pseudo-polar coordinates.

3) With unit Fr\'{e}chet margins, the stability property \eqref{eq:stability}  
can be rephrased by $G^a_0(a\by)=G_0(\by)$ 
for any $a>0$, implying that $\eta$ satisfies the homogeneity property 
\begin{equation}\label{eq:homog}
\eta(a\cB_{\by})=\eta(\cB_{\by})/a,
\end{equation}
for all $\cB_{\by}\subset \overline{\real}$, where 
$\cB_{\by}:=\overline{\real} \backslash(\bzero,\by]$ with $\by>\bzero$. Note that
for a Borel set $\cB\subset \overline{\real}$ we have $a\cB=\{a\bv:\bv\in \cB\}$ and
$\cB_{a\by}=a\cB_{\by}$.
Now, let
$$
\sW:=(\bv\in\overline{\real}: v_1+\ldots+v_d=1),
$$ 
be the unit simplex on $\overline{\real}$ (simplex for brevity), 
where $d-1$ variables are free to vary and one is fixed, e.g.
$v_d=1-(v_1+\cdots+v_{d-1})$. For any $\bv\in \real_+^d$, 
with the sum-norm, $\|\bv\|=|v_1|+\cdots+|v_d|$, we measure the distance of $\bv$ from $\bzero$.
Other norms can also be considered \citep[e.g.][pp.~270--274]{resnick2007}. 
We consider the one-to-one transformation 
$Q: \overline{\real}\rightarrow (0,\infty)\times \sW$, given by 
$$
(r,\bw):=Q(\bv)=(\|\bv\|,\|\bv\|^{-1}\bv),\quad \bv\in\overline{\real}.
$$
By means of this, the induced measure is $\psi:=\eta \ast Q$, i.e. 
$\psi(\cW_{r})=\eta\{Q^{\leftarrow}(\cW_{r})\}$ for all sets $\cW_{r}=r\times \cW$
with $r>0$ and $\cW\subset \sW$, is generated. Then, from the property \eqref{eq:homog}
it follows that 
\begin{equation*}
\begin{split}
\psi(\cW_r)&=\eta\{(\bv\in\overline{\real}:\|\bv\|>r,\bv/\|\bv\|\in \cW)\}\\
&=\eta\{(r\bu\in\overline{\real}:\|\bu\|>1,\bu/\|\bu\|\in \cW)\}\\
&=r^{-1}H'(\cW),
\end{split}
\end{equation*}
where $H'(\cW):=\eta\{(\bu\in\overline{\real}:\|\bu\|>1,\bu/\|\bu\|\in \cW)\}$.
The benefit of transforming the coordinates into pseudo-polar is that the measure $\eta$ becomes a product
of two independent measures: the {\it radial measure} ($1/r$) and {\it spectral measure} or {\it angular measure} 
($H'$) \citep[e.g.][p.~145]{falk2011}.
\index{Angular measure}
The first measures the intensity (or distance) of the points from the origin and the second
measures the angular spread (or direction) of the points. This result is 
known as the {\it spectral decomposition} \citep{dehaan1977}. Hereafter we will use the term angular measure.
\index{Spectral decomposition}

The density of $\psi$ is $\der \psi(r,\bw)=r^{-2} \der r\times \der H'(\bw)$ for all $r>0$ and $\bw\in\sW$, by 
means of which we obtain the explicit form
\begin{equation}\label{eq_trans_set}
\begin{split}
\eta(\cB_{\by})&=\psi\{Q(\bv \in \overline{\real}: v_1>y_1 \, \text{or}\ldots \text{or} \,v_d>y_d)\}\\
&=\psi[\{(r,\bw) \in (0,\infty)\times \sW: r>\min(y_j/w_j,j\in I)\}]\\
&=\int_{\sW}\int_{\min(y_j/w_j, j\in I)}^{\infty}r^{-2} \der r\der H'(\bw)\\
&=\int_{\sW}  \max_{j\in I}\left(w_j/y_j\right) \der H'(\bw).
\end{split}
\end{equation}
In pseudo-polar coordinates, extremes are the values whose radial component is higher than a high threshold, 
see the red points in the middle panels of Figure \ref{fig:extremal_dep}.
The angular components are concentrated around the center of the simplex,
in the case of strong dependence (middle-top panel), while they are concentrated around the vertices of the simplex (middle-bottom panel), in the case of weak dependence.     

The measure $H'$ can be any finite measure on $\sW$ 
satisfying the first moment conditions 
\begin{equation*}\label{eq:moment_condition}
\int_{\sW}w_j \,\der H'(\bw)=1,\quad \forall \, j\in I.
\end{equation*}
This guarantees that the marginal distributions of $G_0$ are unit Fr\'{e}chet.
If $H'$ satisfies the first moment conditions, then the total mass is equal to
\begin{equation*}\label{eq:total}
H' (\sW)= \int_{\sW} ( w_1 + \cdots + w_d )\der H'(\bw)
=\sum_{j\in I} \int_{\sW} w_j \der H'(\bw) = d.
\end{equation*}
So setting 
$H:=H'/H'(\sW)$,
then $H$ is a probability measure satisfying
\begin{align}\label{eq:center_of_mass_cond}
\int_{\sW}w_j \der H(\bw)=1/d,\quad \forall \, j\in I.
\end{align}
Concluding, combining \eqref{eq:limit_fre}, \eqref{eq:exp_fun}, \eqref{eq:prob_maxstab} 
and \eqref{eq_trans_set} all together, we have that
a MEVD with unit Fr\'{e}chet margins is equal to 
\begin{equation}\label{eq:dist_maxstab_simp}
G_0(\by)=\exp
\left\{
-d\int_{\sW}  \max_{j\in I}\left(w_j/y_j\right) \der H(\bw)
\right\}.
\end{equation}
%

\subsection{Angular densities}\label{ss:ang_den}

The measure $H$ can place mass on the interior as well as on
other subspaces of the simplex, such as the edges and the vertices. 
Thus $H$ can have several densities that lie on these sets,
which are named {\it angular densities}.
\index{Angular density}
\citet{coles1991} described 
a way to derive the angular densities when $G$ is absolutely continuous 
\citep[see also][Example.~5.13]{resnick2007}. 

Specifically, let $\sS:=\sP(I)\backslash\varnothing$, where $\sP(I)$ is the power set of $I$ and $\cS$ be the index set that takes values in $\sS$. Given fixed $d$, the sets 
$$
\sW_{d,\cS}=(\bw \in \sW : w_j =0, \text{ if } j \notin \cS;\, w_j>0\text{ if } j\in \cS),
$$
for all $\cS\in \sS$ provide a partition of $\sW$ in $2^d-1$ subsets.
Similar to the simplex,
there are $k-1$ variables $w_j$ in $\sW_{d,\cS}$ that are free to vary,
where $j\in \cS$ and $k=|\cS|$ denotes the size of $\cS$.
We denote by $h_{d,\cS}$ the density that lies on the subspace 
$\sW_{d,\cS}$, where $\cS\in \sS$. When the latter is a vertex $\be_j$ of the simplex $\sW$, 
for any $j\in I$, then the density is a point mass, that is $h_{d,\cS}=H(\{\be_j\})$.

Let $\cS=\{i_1,\ldots,i_k\}\subset I$, when $G_0$ is absolutely continuous the angular density
for any $\by\in\real^d_+$ is
\begin{equation}\label{eq:spectralden}
h_{d,\cS} \left( \frac{y_{i_1}}{\sum_{\substack{i\in \cS}} y_i}, \cdots , \frac{y_{i_{k-1}}}{\sum_{i \in \cS} y_i} \right)=
-\left(\sum_{\substack{i\in \cS}}  y_i \right)^{(k+1)} 
\lim_{\substack{y_j\rightarrow0,\\ j\notin \cS}}
\frac{\partial^k V}{\partial y_{i_1} \cdots \partial y_{i_k}} ( \by ).
\end{equation}
Two examples of a tridimensional angular density in the interior of the simplex  
are reported in the right panels of Figure \ref{fig:extremal_dep}.
These are the densities of a symmetric logistic model \citep{gumbel60} with a strong and weak dependence.
When $\cS=\{i\}$ for any $i \in I$ the angular density $h_{d,\cS}$ represents the mass of $H$ at the vertex $\be_j$ with $j=i$, thus
\eqref{eq:spectralden} reduces into
\begin{equation}\label{eq:spectralden_biv}
h_{d,\cS} =H(\{\be_i\})=
-y_i^{(2)} 
\lim_{y_j\rightarrow0,j\notin \cS}
\frac{\partial V}{\partial y_i} ( \by ).
\end{equation}

In the bivariate case these results are equal to the ones obtained by \citet{pickands1981}.
\citet{kotz2000} discussed the bivariate case in the following terms. 
With $d=2$ the unit simplex $\sW = [0,1]$ can be
partitioned into 
$$
\sW_{2,\{1\}}=\{(1,0)\},\quad \sW_{2,\{2\}}=\{(0,1)\},\quad \sW_{2,\{1,2\}}=\{(w,1-w),w\in (0,1)\}.
$$
The densities that lie on them are
$$
h_{2,\{1\}}=H(\{0\})=-y_1^2\lim_{y_2\rightarrow 0}\frac{\partial V}{\partial y_1}(y_1,y_2),
$$
$$
h_{2,\{2\}}=H(\{1\})=-y_2^2\lim_{y_1\rightarrow 0}\frac{\partial V}{\partial y_2}(y_1,y_2),
$$
and
$$
h_{2,\{1,2\}}(w)=-\frac{\partial^2 V}{\partial y_1\,\partial y_2}(w,1-w).
$$
respectively, for any $y_1,y_2>0$.
The first two densities describe the case when extremes are only observed in one variable. 
While the third density describes the case when extremes are observed in both variables.

\subsection{Extremal dependence}\label{ss:dep}

From \eqref{eq:prob_maxstab} it emerges that the extremal dependence is expressed through the
exponent function.
\index{Extremal dependence}
This is a map from $\real^d_+$ to $(0,\infty)$ satisfying the properties:
 \begin{enumerate}
\item is a continuous function and homogeneous of order $-1$, the latter meaning that $V(a\by)=a^{-1}V(\by)$
for all $a>0$;
\item is a convex function, that is $V(a\by+(1-a)\by')\leq aV(\by)+(1-a)V(\by')$,
for $a\in[0,1]$ and $\by,\by'\in \real^d_+$;
\item
$
\max\left(1/y_1,\ldots,1/y_d\right) \leq V(\by)\leq (1/y_1+\ldots+1/y_d),
$
with the lower and upper limits representing the complete dependence and independence cases respectively.
\end{enumerate}
See \citet[pp.~223--226]{dehaan2006} for details. In summary, let $\bY$ be a random vector with
distribution \eqref{eq:dist_maxstab_simp}. When $H$ places the total mass 1 on 
the center of the simplex $(1/d,\ldots,1/d)$, then $Y_1=Y_2=\cdots=Y_d$ almost surely and 
hence $G_0(\by)=\exp\{\max\left(1/y_1,\ldots,1/y_d\right)\}$.
When $H$ places mass $1/d$ on $\be_j$ for all $j\in I$, i.e. the vertices of the simplex, then
$Y_1,\ldots,Y_d$ are independent and hence  
$G_0(\by)=\exp(1/y_1+\ldots+1/y_d)$.
This rephrased for a random vector $\bX$ with distribution
\eqref{eq:limMEVD} becomes
$$
\min\{G_1 (x_1),\ldots,G_d (x_d)\} \leq G (\bx)\leq G_1( x_1)\cdot \ldots \cdot G_d( x_d), \quad \bx\in \real^d.
$$

In order to visualise the exponent function more easily, its restriction in the simplex is usually considered. 
\index{Pickands dependence}
This is a function $A: \sW\rightarrow [1/d,1]$, named the {\it Pickands dependence} function
\citep{pickands1981}, defined by
$$
A(\bt):=d\int_{\sW}\max_{j\in I}
\left(
w_j\,t_j
\right)\der H(\bw),
$$
where $z_j=1/y_j$, $j\in I$, $t_j=z_j/(z_1+\cdots+z_d)$ with $j=1,\ldots,d-1$ and $t_d=1-(t_1+\cdots+t_{d-1})$. $A$ inherits the above 
properties from $V$ with the obvious modifications.
In particular, $1/d\leq\max(t_1,\ldots,t_d)\leq A(\bt)\leq1$, where lower and upper bounds represent the
complete dependence and independence cases, and 
for the homogeneity property of $A$ the exponent function can be rewritten as
\begin{equation*}\label{eq:exp_pick}
V(\bz)=(z_1+\cdots+z_d)A(t_1,\ldots,t_d),\quad \bz\in \real^d_+.
\end{equation*}

The exponential function can be profitably used in several ways. First, an important summary 
of the extremal dependence is given by
\begin{equation}\label{eq:extremalcoeff}
\vartheta=V(1,\ldots,1)=d\int_{\sW} \max_{j\in I} (w_j) \der H(\bw).
\end{equation}
This is named the {\it extremal coefficient} \citep{smith1990a} and it
represents the (fractional) number of independent components of the random vector $\bY$.
\index{Extremal coefficient}
The coefficient takes values in $[1,d]$, depending on whether the measure $H$ concentrates near the center or the vertices of the simplex. 
The bounds regard the cases of complete dependence and independence.

Second, for any $\by>\bzero$ and failure region 
\begin{equation}\label{eq:failure}
\cF_{\by}=(\bv \in \overline{\real}: v_1>y_1 \text{ and} \ldots \text{and } v_d>y_d),
\end{equation}
the {\it tail dependence} function \citep[][p.~225]{nikoloulopoulos2009, dehaan2006} 
is defined by
\begin{equation*}\label{eq:taildep}
R(\by):=\eta\{(\bv \in \overline{\real}: v_1>y_1 \text{ and} \ldots \text{and } v_d>y_d)\}\equiv \eta(\cF_{\by}),
\quad \by>\bzero.
\end{equation*}
\index{Tail dependence function}
This counts the number of observations that fall in the failure region, i.e. all their coordinates are simultaneously large. 
The tail dependence function is related to the exponent function by the inclusion-exclusion principle.
Using similar arguments to those in \eqref{eq_trans_set} and \eqref{eq:center_of_mass_cond} it follows that 
\begin{equation}\label{eq:tail_measure}
R(\by)=d\int_{\sW} \min_{j\in I}(w_j/y_j) \der H(\bw)
\quad \by>\bzero.
\end{equation}
By means of the tail dependence function, another 
important summary of the dependence between the components of $\bY$
is obtained.
The {\it coefficient of upper tail dependence} is given by
\begin{equation}\label{eq:taildep_coef}
\chi=R(1,\ldots,1)=d\int_{\sW} \min_{j\in I} (w_j) \der H(\bw).
\end{equation}
It measures the strength of dependence in the tail of the distribution of $\bY$ or in other
terms the probability that all the components of $\bY$ are simultaneously large.
This coefficient was introduced in the bivariate case by  \citet[][Ch.~2]{joe1997} and 
extended to the multivariate case by \citet{li2009orthant}.
When $H$ concentrates near the center or on the vertices of the simplex, then $\chi>0$ or 
$\chi=0$ respectively. In these cases we say 
that $\bY$ is upper tail dependent or independent.
\index{Coefficient of upper tail dependence}

In addition, the exponent and the tail dependence functions can be used for approximating the
probability that certain types of extreme events will occur. Specifically, let $\bY$ be a random vector
with unit Pareto margins. $F$ is in the domain of attraction of a MEVD with Fr\'{e}chet margins. From 
\eqref{eq:exp_fun} and for the homogeneity property of $V$ we have that $\{1-F(n\by)\}\approx V(n\by)$ for large $n$.
Then, for the relations \eqref{eq_trans_set} and \eqref{eq:center_of_mass_cond}, the approximating result follows
\begin{equation}\label{eq:almost_fail_proxy}
\P(Y_1>y_1\text{ or } \ldots \text{ or } Y_d>y_d)\approx d\int_{\sW} \max_{j\in I} \left(w_j/y_j\right) \der H(\bw),
\end{equation}
when $y_1,\ldots,y_d$ are high enough thresholds. 
Furthermore, with similar arguments to those 
in Section \ref{ss:maxstab} we have that
$$
\lim_{n\rightarrow\infty}n\bar{F}(ny_1,\ldots, ny_d)=R(\by),
$$
where $\bar{F}$ is the survivor function of $\bY$. $R$ has the same homogeneity property of $V$.
Hence, $\bar{F}(n\by)\approx R(n\by)$ for large $n$.  Then,
for the relation \eqref{eq:tail_measure}, the approximating result also follows
\begin{equation}\label{eq:prob_failure}
\P(Y_1>y_1 \text{ and} \ldots \text{and } Y_d>y_d)\approx 
d\int_{\sW} \min_{j\in I} \left(w_j/y_j\right) \der H(\bw),
\end{equation}
when $y_1,\ldots,y_d$ are high enough thresholds.

Lastly, when $\chi=0$ the elements of $\bY$ are independent in the limit. However, they may still be dependent for large but finite samples. 
\citet{ledford+t1996} proposed another dependence measure in order to capture this feature. 
For brevity, we focus on the bivariate case. Suppose that $\bar{F}$ for $y\rightarrow\infty$ satisfies the condition
$$
\bar{F}(y,y)\approx y^{-1/\tau}\cL(y),\quad 0<\tau \leq 1,
$$
where $\cL$ is a slowly function, i.e. $\cL(ay)/\cL(y)\rightarrow 1$ as $y\rightarrow\infty$ for any $a>0$. Then for large $y$, assuming
$\cL$ constant, different tail behaviours are covered. 
The case $\chi>0$ is reached when $\tau=1$  and so the variables are asymptotically dependent. 
When $1/2<\tau<1$ this means that $\chi=0$ and so the variables are asymptotically independent, but they are still  positively associated
and the value of $\tau$  expresses the degree \citep[see][for details]{ledford+t1996}.

\section{Parametric models for the extremal dependence}\label{s:par_models}

From the previous sections, it emerges that both the exponent and tail dependence functions
depend on the angular measure.
There is no unique angular measure that generates the extremal dependence, 
any finite measure that satisfies the first moment conditions is suitable.
In order to represent the extremal dependence, in principle it is insufficient to
use a parametric family of models for the distribution function of the angular measure.
However, flexible classes of parametric models can still be useful for applications, e.g. see \citet{tawn1990}, \citet{coles1991} and \citet{boldi2007} to name a few.
To this end, in previous years different parametric extremal dependence models
have been introduced in the literature. A fairly comprehensive overview can be found in
\citet[][Section 3.4]{kotz2000}, \citet[][Section 8.2.1]{coles2001}, \citet[][Section 9.2.2]{beirlant2006} and \citet{padoan2013b}.
In the next sections we describe some of the most popular models.

\subsection{Asymmetric logistic model}\label{sec:asym}

The multivariate asymmetric logistic model is an extension of the symmetric,
introduced by \citet{tawn1990} \citep[see also][]{coles1991}
for modelling extremes in complex environmental applications.
\index{Asymmetric logistic model}

Let $\sS$ and $\cS$ as in Section \ref{ss:ang_den} and
$N_\cS$ be a Poisson random variable with rate $1/\tau_\cS$.
This describes the number of storm events, $n_\cS$, that takes place on the sites $\cS$
in a time interval.
Given $n_\cS$, for any site $j\in \cS$, let $\{X_{j,\cS;i},i=1,\ldots,n_\cS\}$ be a sequence of i.i.d. random variables that describe an environmental episode such as rain. 
For a fixed $i$, $\{X_{j,\cS;i}\}_{j\in \cS}$ is assumed to be a dependent sequence.
The maximum amount of rain observed at $j$ is $X_{j,\cS}=\max_{i=1\ldots,n_\cS}\{X_{j,\cS;i}\}$.
Let $A_\cS$ be a random effect with a positive stable distribution and stability parameter 
$\alpha_\cS \geq 1$ \citep{nolan2003}, representing an unrecorded additional piece of information on storm events. 
Assume $\{X_{j,\cS}\}_{j\in \cS}|\alpha_\cS$ as an independent sequence. Define
$
Y_j=\max_{\cS\in \sS_j}\{X_{j,\cS}\},
$
where $\sS_j\subset \sS$ contains all nonempty sets including $j$ and so the maximum is over all the storm
events involving $j$. Then, the exponent function of the joint survival function of $(Y_1,\ldots,Y_d)$, after 
transforming the margins into unit exponential variables, is 
\begin{equation*}\label{eq:exponent_logistic}
V(\by;\btheta)=\sum_{\cS\in \sS} \Big\lbrace \sum_{j \in \cS} ( \beta_{j,\cS} y^{-1}_j )^{\alpha_\cS} \Big\rbrace^{1/\alpha_\cS},\quad \by \in \real^d_+,
\end{equation*}
where $\btheta=\{\alpha_\cS,\beta_{j,\cS}\}_{\cS\in \sS}$, 
$ \alpha_\cS \geq 1$, $\beta_\cS=\tau_\cS/\sum_{\cS\in \sS_j} \tau_\cS$ and 
$\beta_{j,\cS} = 0$ if $j\notin \cS$,
and for $j \in I$, $0\leq \beta_{j,\cS}\leq 1$ and $\sum_{\cS\in \sS} \beta_{j,\cS}=1$. 
The parameter $\beta_{j,\cS}$ represents the probability that the maximum value observed at $j$
is attributed to a storm event involving the sites of $\cS$.
The number of the model parameters is $2^{d-1} (d+2) - (2d+1)$.

In this case the angular measure places mass on all the subspaces of the simplex. 
From \eqref{eq:spectralden} it follows that the angular density is, for every $\cS\in \sS$ 
and all $\bw\in \sW_{d,\cS}$ equal to
\begin{equation*}\label{eq:asm_specden}
h_{d,\cS} ( \bw;\btheta ) = 
\prod_{i=1}^{k-1} ( i\alpha_\cS-1 )
  \prod_{j \in \cS} \beta_{j,\cS}^{\alpha_\cS} w_{j}^{- (\alpha_\cS+1)}    
\Big\lbrace \sum_{j \in \cS} (\beta_{j,\cS}/ w_j)^{\alpha_\cS} \Big\rbrace^{1/\alpha_\cS-k}. 
\end{equation*} 
When $\cS=I$,
$\alpha_\cS=\alpha$, $\beta_{j,\cS}=\beta_j$ and so
the angular density on the interior of the simplex simplifies to
\begin{equation*}
h ( \bw ;\btheta) = \prod_{i=1}^{d-1} ( i\alpha -1)
  \prod_{j\in I} \beta_{j}^{\alpha} w_{j}^{- (\alpha+1)}  
\Big\lbrace \sum_{j\in I} (\beta_{j}/w_j)^{\alpha} \Big\rbrace^{1/\alpha-d},\quad \bw \in \sW.
\end{equation*}
When $\cS=\{j\}$, for all $j\in I$, then from \eqref{eq:spectralden_biv} it follows that the
point mass at each extreme point of the simplex is $h_{d,\cS}=\beta_{j,s}$.

For example in the bivariate case, the conditions on the parameters are $\beta_{1,\lbrace 1 \rbrace} + \beta_{1,\lbrace 1,2 \rbrace} = 1$ and $\beta_{2,\lbrace 2 \rbrace} + \beta_{2,\lbrace 1,2 \rbrace} = 1$,
so the masses at the corners of $\sS_2 = [0,1]$ are given by 
$ h_{2,\lbrace 1 \rbrace } = 1 - \beta_1 $ and $ h_{2,\lbrace 2 \rbrace } = 1 - \beta_2 $, where for simplicity
$\beta_{1,\lbrace 1,2 \rbrace} = \beta_1$ and $\beta_{2,\lbrace 1,2 \rbrace} = \beta_2$, 
while the density in the interior of the simplex, for $0< w <1$, is
$$
h_{2,\lbrace 1,2 \rbrace } (w)
= ( \alpha -1 ) ( \beta_1 \beta_2 )^{\alpha} \lbrace w ( 1-w ) \rbrace^{\alpha -2}
 [ ( \beta_1 ( 1-w ) )^\alpha + ( \beta_2 w )^\alpha]^{1/\alpha -2}.
$$
\begin{figure}[h!]
\begin{center}$
\begin{array}{c}
\includegraphics[width=0.95\textwidth]{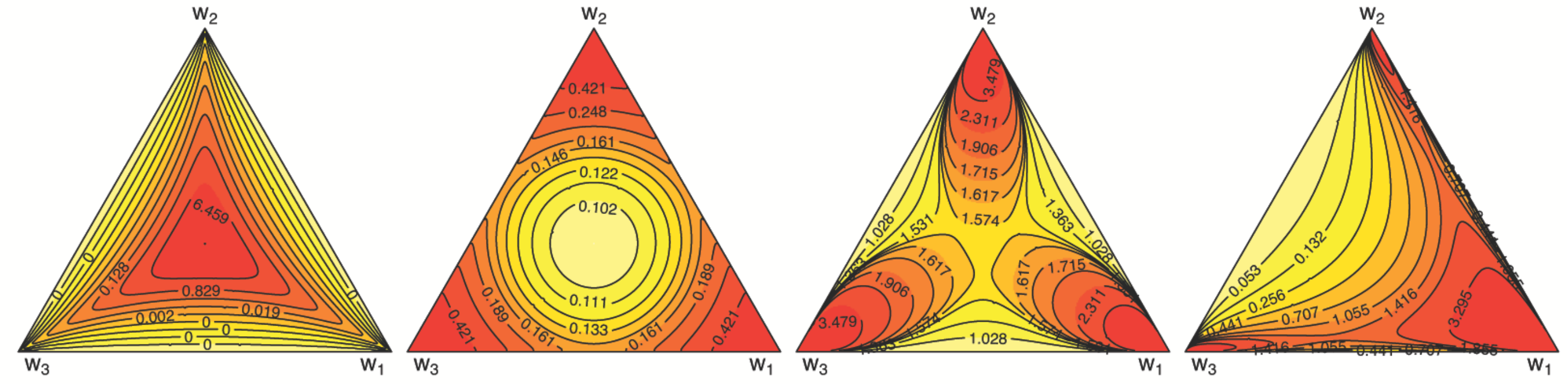} \\
\includegraphics[width=0.95\textwidth]{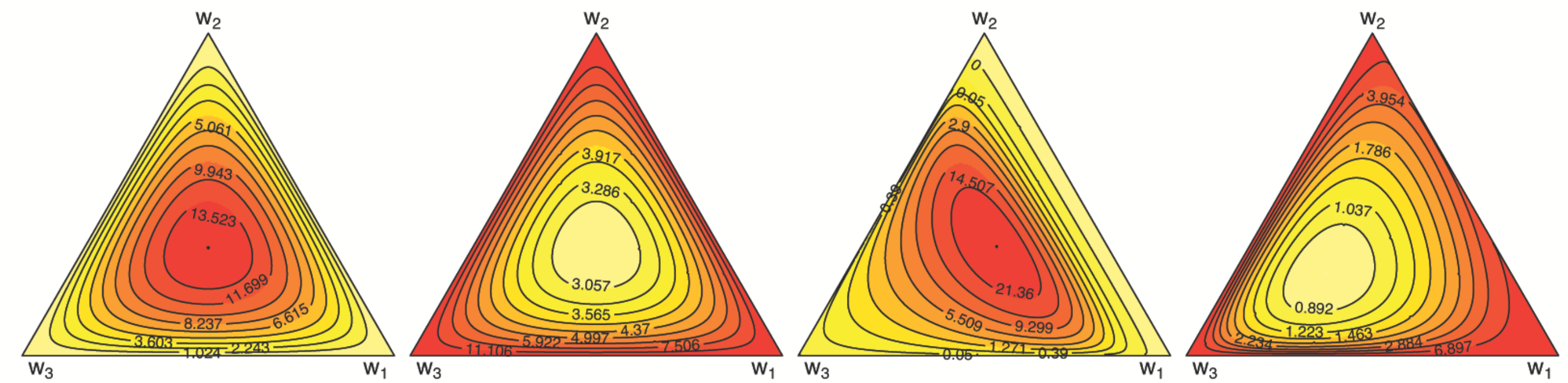} \\
\includegraphics[width=0.95\textwidth]{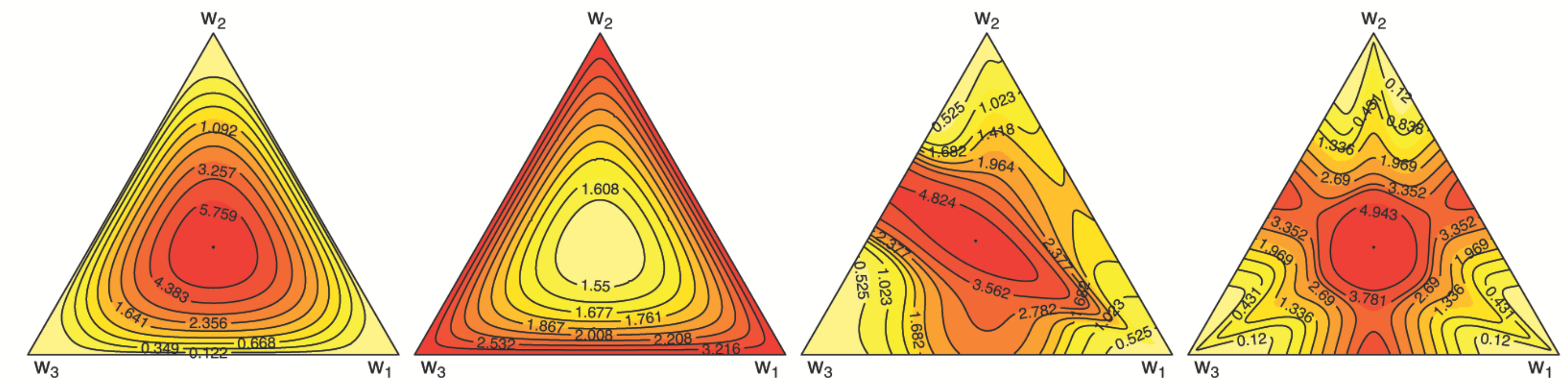} \\
\includegraphics[width=0.95\textwidth]{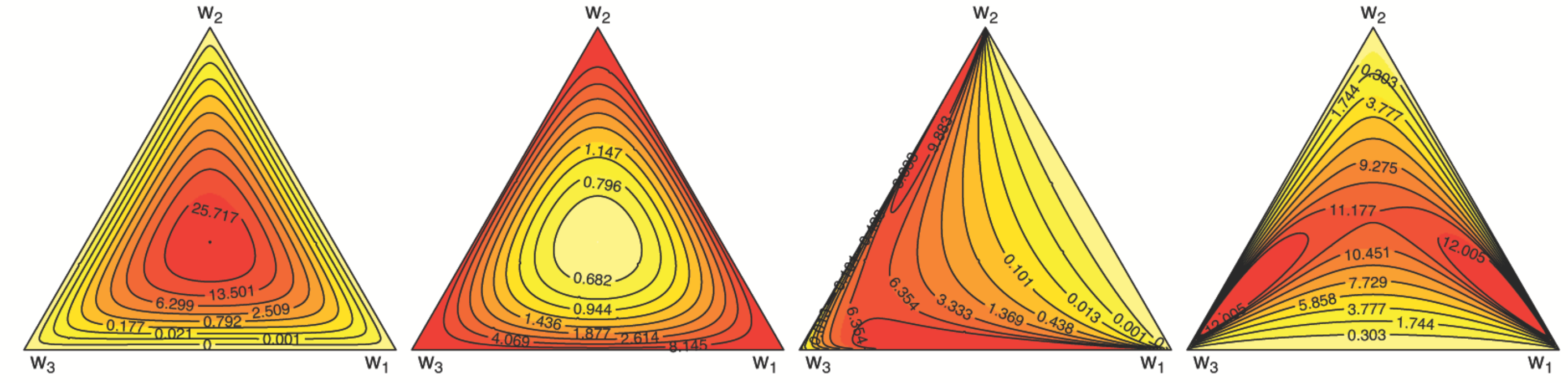} \\
\includegraphics[width=0.95\textwidth]{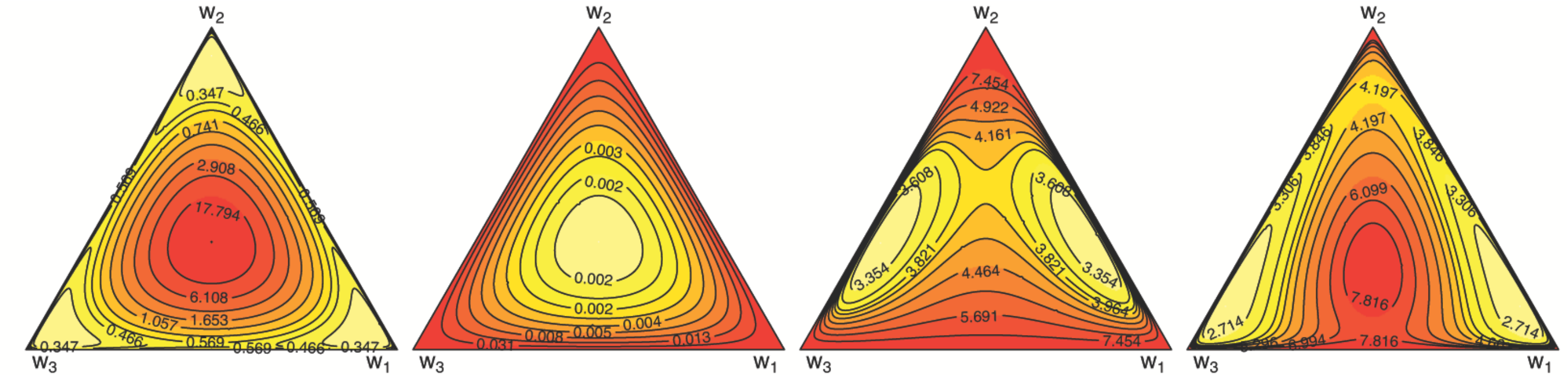}
\end{array}$
\end{center}
\caption
[Examples of parametric angular density models]
{Examples of trivariate angular densities for the Asymmetric Logistic, Tilted Dirichlet,
Pairwise Beta, H\"{u}sler-Reiss and Extremal-$t$ models from top to bottom.}
\label{fig:specden_plot}
\end{figure}
The top row of Figure \ref{fig:specden_plot} illustrates some examples of trivariate angular densities 
for different values of the parameters $\btheta = ( \alpha,\beta_1,\beta_2,\beta_3 )$, where the subscript of the index set 
$\cS = \lbrace 1,2,3 \rbrace$ has been omitted for simplicity.
The values of the parameters are, from left to right 
$\{( 5.75, 0.5, 0.5, 0.5 );( 1.01, 0.9, 0.9, 0.9 );( 1.25, 0.5, 0.5, 0.5 );( 1.4, 0.7, 0.15, 0.15 )\}$. 
The first panel shows that with large values of $\alpha$ and equal values of the other parameters, the case of strong dependence among the variables is obtained. The mass is mainly concentrated towards the center of the simplex.
The second panel shows that when $\alpha$ is close to $1$ and the other parameters are equal, the
case of weak dependence is obtained. The mass is concentrated on the vertices of the simplex.
The third panel shows the case of a symmetric dependence structure with the mass near the corners of the simplex but not along the edges. Finally, the fourth panel shows a case of an asymmetric dependence structure where
 the mass tends to be closer to the components whose corresponding values of $\beta$ are high.

\subsection{Tilted Dirichlet model}

Extremal dependence models with an angular measure that places mass on the interior, vertices and edges of the simplex
are more flexible than those with a measure that concentrates only on the interior.
An example is the asymmetric logistic model versus the symmetric.
However, the former has too many parameters to estimate, so parsimonious models may be preferred.
\index{Tilted Dirichlet model}
In order to derive a parametric model for the angular density whose mass concentrates on the interior of the simplex, 
\citet{coles1991} proposed the following method.
Consider a continuous function $h': \sW \rightarrow [0,\infty)$ such that
$
m_j = \int_{\sS_d} v_j \,h'( \bv ) \mathrm{d}\bv <\infty
$
for all $j\in I$. Then, the function
\begin{equation*}\label{eq:spectral_transformation}
h ( \bw ) = d^{-1}(m_1w_1+\cdots+m_dw_d)^{-(d+1)}
h' \{\bm \bw /(m_1w_1+\cdots+m_dw_d)\},\bw \in \sW
\end{equation*}
is a valid angular density.
It satisfies the first moment conditions \eqref{eq:center_of_mass_cond}
and its mass is centered
at $(1/d,\ldots,1/d)$ and integrates to one.
For example, if $h'$ is the density of the Dirichlet distribution, then
we obtain the angular density 
\begin{align} \label{eq:density_dirichlet}
h ( \bw ;\btheta)= 
\frac{\Gamma (\sum_{j\in I}\alpha_j+1 )}
{d(\sum_{j\in I}\alpha_jw_j)^{d+1}}
\prod_{j=1}^d \frac{\alpha_j}{\Gamma ( \alpha_j )}
\left( \frac{\alpha_j w_j}{\sum_{j\in I}\alpha_jw_j} \right)^{\alpha_j - 1},\bw\in \sW,
\end{align}
where $\btheta=\{\alpha_j>0\}_{j\in I}$.
This density is asymmetric and it becomes symmetric when
$\alpha_1 = \cdots = \alpha_d$. 
Extremes are independent or completely dependent when for all $j\in I$ the limiting cases
$ \alpha_j \rightarrow 0$ and $ \alpha_j \rightarrow \infty$ arise.
The dependence parameters $\alpha_j$, $j\in I$, are not easy to interpret. 
However, \citet{coles1994} draw attention to the quantities
$r_1 = (\alpha_i - \alpha_j)/2$ and $r_2 = (\alpha_i + \alpha_j)/2$ which can be interpreted as 
the asymmetry and intensity of the dependence between pairs of variables.

In this case, the exponent function can not be analytically computed, nonetheless it can still be evaluated numerically.

The second row of Figure \ref{fig:specden_plot} illustrates some examples of trivariate angular densities obtained with different sets of the parameters $\btheta = ( \alpha_1, \alpha_2, \alpha_3 )$.
The plots from left to right have been obtained using the parameter sets
$\lbrace ( 2, 2, 2 ); ( 0.5, 0.5, 0.5 ); ( 2, 2.5, 30 ); ( 0.1, 0.25, 0.95 ) \rbrace $.
The first panel shows that, when values of the parameters are equal and greater than $1$, the mass concentrates
in the center of the simplex leading to strong dependence. 
The second panel shows the opposite, when values of $\boldsymbol \alpha$ are equal and less than $1$, it yields to the case of weak dependence as the mass concentrates on the vertices of the simplex.
The third panel shows the case of an asymmetric dependence structure and this is obtained when the values of the parameters are all greater than one.
In this specific case the mass tends to spread towards the bottom and top left edges.
The fourth panel illustrates another case of an asymmetric dependence structure, in this case obtained with
all the values of the parameters that are less than $1$, leading to a mass that concentrates along the top right edge and vertices.   

\subsection{Pairwise beta model}

The tilted Dirichlet model has been successfully used for applications \citep[e.g.][]{coles1991}, 
although it suffers from a lack of interpretability of the parameters. \citet{cooley2010} proposed
a similar model but with easily interpretable parameters. The definition of their model is based on a geometric approach.
\index{Pairwise beta model}
Specifically, they considered 
the symmetric pairwise beta function
\begin{align*}
h^*(w_i,w_j) = \frac{\Gamma( 2 \beta_{i,j} )}{\Gamma^2( \beta_{i,j} )}
\bigg( \frac{w_i}{w_i + w_j} \bigg)^{\beta_{i,j}-1}
\bigg( \frac{w_j}{w_i + w_j} \bigg)^{\beta_{i,j}-1},i,j\in I,
\end{align*} 
where $w_i$ and $w_j$ are two elements of $\bw$ and $\beta_{i,j}>0$.
This function has its center at the point $(1/d,\ldots,1/d)$ and it
verifies the first moment conditions \eqref{eq:center_of_mass_cond}.  
Then, the angular pairwise beta density is defined by summing together all the $d(d-1)/2$ possible pairs of 
variables, namely
\begin{align*}
h ( \bw;\btheta) = \frac{2 ( d-3 )!\Gamma ( \alpha d +1 )}{d (d-1)\Gamma ( 2\alpha +1 )\Gamma \{ \alpha (d-2) \}}
\sum_{i,j\in I,i<j} 
h(w_i,w_j), \quad \bw \in \sW,
\end{align*}
where
$$
h(w_i,w_j)=(w_i+w_j)^{2\alpha-1}\{1-(w_i+w_j)\}^{\alpha(d-2)-d+2}\,h^*( w_i,w_j)
$$
and $\btheta=(\alpha,\{\beta_{i,j}\}_{i,j\in I})$ with $\alpha>0$.
Each parameter $\beta_{i,j}$ controls the level of dependence between the $i^{th}$ and the $j^{th}$ components
and the dependence increases for increasing values of $\beta_{i,j}$.
The function $h^*$ is introduced to guarantee that the dependence 
ranges between weak and strong dependence.
The parameter $\alpha$ controls the dependence of all the variables, when it increases
the overall dependence increases.

Also in this case the exponent function can not be computed in closed form and hence
it can only be evaluated numerically.

The third row of Figure \ref{fig:specden_plot} provides some examples of trivariate angular densities obtained with different values of the parameters 
$ \btheta = (\alpha, \beta_{1,2}, \beta_{1,3}, \beta_{2,3})$.
The plots from left to right have been obtained using the parameter sets 
$
\lbrace (4, 2, 2, 2); (0.5, 1, 1, 1); (1, 2, 4, 15);(1, 10, 10, 10) \rbrace $.
The first panel shows a case of symmetric density obtained with all equal parameters $\beta_{i,j}$ $i,j\in I$. A large value of the overall dependence parameter $\alpha$ pulls the mass towards the center of the simplex, indicating a strong dependence between the variables.
On the contrary, the second panel shows that when the overall dependence parameter is close to zero then the mass concentrates on the vertices of the simplex, indicating weak dependence among the variables.
The third panel illustrates a case of asymmetric angular density with strong dependence between the second and third variables that is due 
to a large value of $\beta_{2,3}$.
Although the value of the global dependence parameter $\alpha$ is not large, it is enough to slightly push the mass towards the center of the simplex.
The fourth panel shows a case of symmetric angular density, which is obtained
with large values of the pairwise dependence parameters and an average value of the global dependence parameter.
The mass is mainly concentrated on the center of the simplex and some mass tends to lie near the centers of the edges.

\subsection{H\"{u}sler-Reiss model}

One of the most popular models is the H\"{u}sler-Reiss \citep{husler1989}.
\index{H\"{u}sler-Reiss model}
Let $\bX_1,\ldots,\bX_n$ be $n$ i.i.d. copies of a zero-mean unit variance Gaussian random vector. Assume that for all $i,j\in I$  
the pairwise correlation $\rho_{i,j;n}$ satisfies the condition 
$$
\lim_{n\rightarrow \infty} \log n(1-\rho_{i,j;n})=\lambda_{i,j}^2 \in [0,\infty).
$$
Then, the exponent function of the limit distribution of $\bb_n(\bM_n-\bb_n)$ for $n\rightarrow \infty$, 
where $\bb_n=(b_n,\ldots,b_n)$ is a vector of real sequences \citep[see][pp. 71-72]{resnick2007},
is 
\begin{equation}\label{eq:exponent_husler}
V ( \by ;\btheta) = \sum_{j=1}^d  \frac{1}{y_j} \Phi_{d-1} 
\Big\lbrace \Big ( \lambda_{i,j} +\frac{\log y_i/y_j }{2\lambda_{i,j}} \Big )_{i \in I_j}; \bar{\Lambda}_j \Big\rbrace, 
\quad \by \in \real^d_+,
\end{equation}
where $\btheta=\{\lambda_{i,j}\}_{i,j\in I}$, $I_j := I\setminus\lbrace j \rbrace$, $\Phi_{d-1}$ is $d-1$ dimensional Gaussian distribution with
partial correlation $\bar{\Lambda}_j$. For all $j\in I$, the elements of $\bar{\Lambda}_j$ are $\lambda_{k,i; j} = (\lambda_{k,j}^2 + \lambda_{i,j}^2 - \lambda_{k,i}^2 ) / ( 2 \lambda_{k,j}\lambda_{i,j} )$, for $k,i \in I_j$.
The parameter $\lambda_{i,j}$, $i,j \in I$, 
controls the dependence between the $i^{th}$ and $j^{th}$ elements of a
vector of $d$ extremes. These are completely dependent
when $\lambda_{ij} = 0$ and become independent as
$\lambda_{ij}\rightarrow\infty$. 

In this case the angular measure concentrates on the interior of the simplex. 
Applying \eqref{eq:spectralden}
it can be checked \citep{engelke2012} that the angular density is
\begin{align*}\label{dens_husler}
h( \bw ;\btheta) = \phi_{d-1} \left\{\left(\lambda_{i,1}+\frac{ \log w_i / w_1 }{ 2 \lambda_{i,1} } 
\right)_{i\in I_1}; \bar{\Lambda}_1 \right\}
\left\{w_1^2 \prod_{i=2}^d  ( w_i 2 \lambda_{i,1})\right\}^{-1},\bw \in \sW, 
\end{align*}
where $\phi_{d-1}$ is $d-1$ dimensional Gaussian density with
partial correlation matrix $\bar{\Lambda}_1$.

The second last row of Figure \ref{fig:specden_plot} provides some examples of trivariate angular densities 
obtained with different values of the parameters $\btheta = ( \lambda_{1,2}, \lambda_{1,3}, \lambda_{2,3} )$.
The plots from left to right have been obtained using the parameter sets 
$\{( 0.3, 0.3, 0.3 ), ( 1.4, 1.4, 1.4 ), ( 1.7, 0.7, 1.1 ), ( 0.52, 0.71, 0.52 )\}$.
The first panel shows that with small and equal values of parameters the case of strong dependence among all
the variables is obtained. In this case the mass concentrates around the center of the simplex. On the contrary,
the second panel shows that with large and equal values of the parameters the case of weak dependence is obtained.
In this case the mass is placed close to the vertices of the simplex. The third panel shows that an asymmetric dependence
structure is obtained when the parameter values are different. In this case the mass tends to concentrate around the
vertices and edges that are concerned with the smaller values of the parameters. The fourth panel shows that a symmetric
dependence structure, with respect to the second component is obtained setting the values of two parameters to be equal. 
In this case the mass is equally divided up towards the two vertices and edges that are concerned with the smaller values of the parameters.

\subsection{Extremal-$t$ model}\label{lst:extremalt_model}

The extremal-$t$ model \citep{nikoloulopoulos2009} is more flexible than the H\"{u}sler-Reiss
but it is still simple enough. 
\index{Extremal-$t$ model}
It is easily interpretable and useful in practical applications \citep[see][]{davison2012b}.
Let $\bX_1,\ldots,\bX_n$ be $n$ i.i.d. copies of a zero-center unit scale Student-$t$ random vector with dispersion matrix
$\Sigma$ and $\nu>0$ degrees of freedom (d.f.). Then, 
the exponent function of the limiting distribution of $\bM_n/\ba_n$ for $n\rightarrow\infty$, where $\ba_n=(a_n\ldots,a_n)$ is a vector of positive sequences \citep[see][]{demarta2005}, is
\begin{equation} \label{eq:exponent_extremalt}
V ( \by ;\btheta) = \sum_{j=1}^d  \frac{1}{y_j} T_{d-1, \nu+1} 
\Bigg \lbrace \left[ \sqrt{ \frac{\nu+1}{1-\rho_{i,j}^2 } }
\left\{ \Big( \frac{y_i}{y_j} \Big )^{\frac{1}{\nu}} -\rho_{i,j} \right\} \right]_{i \in I_j}; \bar{\Sigma}_j
 \Bigg\rbrace,
\end{equation}
for all $\by \in \real^d_+$, where $\btheta=(\{\rho_{i,j}\}_{i,j\in I},\nu)$ and $T_{d-1, \nu+1}$ is a $d-1$ dimensional Student-$t$ distribution with $\nu+1$ d.f. 
and partial correlation matrix $\bar{\Sigma}_j$. 
The correlation parameter $\rho_{i,j}$, $i,j \in I$, drives the dependence between
pairs of variables with the dependence that increases with the increasing of $\rho_{i,j}$. 
The parameter $\nu$ controls the overall dependence among all the variables. For decreasing
values of $\nu$ the dependence increases and vice versa.

The H\"{u}sler-Reiss model is a special case of the extremal-$t$. 
Indeed, for all $i,j\in I$ if the correlation parameters of the extremal-$t$ distribution
are equal to $\rho_{i,j;\nu}=1-\lambda^2_{i,j}/\nu$, then this distribution
converges weakly, as $\nu\rightarrow\infty$, to the H\"{u}sler-Reiss \citep[see][]{nikoloulopoulos2009}.

In this case the angular measure places mass on all the subspaces of the simplex. 
When $\cS=I$, then applying \eqref{eq:spectralden} we obtain that the angular density is
\begin{equation*} \label{eq:density_extremalt}
h ( \bw ;\btheta) = 
\frac{t_{d-1,\nu+1} 
\left( \left[ \sqrt{ \frac{\nu+1}{1-\rho_{i,1}^2 } }
\left\{  \left(w_i/w_1\right)^{1/\nu} -\rho_{i,1} \right\} \right]_{i\in I_1} ; \bar{\Sigma}_1
 \right)}
{\nu^{d-1} w_1^{d+1} 
\left\{\prod_{i=2}^d \sqrt{ \frac{\nu+1}{1-\rho_{i,1}^2 } }\left( w_i/w_1\right)^{(\nu -1)/\nu}\right\}^{-1}},
\quad \bw\in\sW,
\end{equation*}
where $t_{d-1,\nu+1}$ is $d-1$ dimensional Student-$t$ density with partial correlation matrix 
$\bar{\Sigma}_1$ \citep[e.g.][]{ribatet13}.  
When
$\cS = \lbrace j \rbrace$, then applying \eqref{eq:spectralden_biv} 
we obtain that the mass on the extreme points of the simplex is
\begin{equation*}\label{eq:mass_extremalt}
h_{d, \cS} = T_{d-1, \nu+1} 
\left[ 
\left\{ -\rho_{i,j}(\nu+1)^{1/2}/(1-\rho_{i,j}^2)^{1/2}
\right\}_{i \in I_j}; \bar{\Sigma}_j
\right], \quad j\in I.
\end{equation*}

The last row of Figure \ref{fig:specden_plot} provides some examples of the trivariate angular densities obtained with different values of the parameters 
$\btheta=( \rho_{1,2}, \rho_{1,3}, \rho_{2,3}, \nu )$.
From left to right the plots are obtained using the parameter values
$\{( 0.95, 0.95, 0.95, 2 ); ( -0.3, -0.3, -0.3, 5); ( 0.52, 0.71, 0.52, 3);$ 
$( 0.52, 0.71, 0.52, 2)\}$.
The first panel shows that when the scale parameters $\rho_{ij}$ are all equal and close to one and the d.f. $\nu$ are small, then the mass concentrates around the center of the simplex and therefore the dependence is strong. 
The second panel shows the opposite, when the correlations are close to zero and the d.f. are high, the mass concentrates around the vertices of the simplex and hence the dependence is weak. 
The third panel shows that when two scale parameters are equal then the dependence structure is symmetric with respect to the second component and the mass tends to concentrate on the top
vertex and the bottom edge and vertices. 
The fourth panel shows that with the same setting but with smaller d.f. the mass is pushed towards the center of the simplex and hence the dependence is stronger.
%

\section{Estimating the extremal dependence}\label{sec:estimation}

Several inferential methods have been explored for inferring the extremal dependence.
Nonparametric and parametric approaches are available. In the first case recent advances are
\citet{gudend+s11}, \citet{gudend+s12} and \citet{marcon2014}, see also the references therein.
Both likelihood based and Bayesian inferential methods have been widely investigated. 
Examples of likelihood based methods are the approximate
likelihood \citep[e.g.][]{coles1994, cooley2010,engelke2012} and the composite likelihood \citep[e.g.][]{padoan2010,davison2012c}. 
Examples of Bayesian techniques are \citet{apputhrai+s11},
\citet{sabourin2013}, \citet{sabourin+n14}.

For comparison purposes in the next section the real data analysis is performed using 
the maximum approximate likelihood estimation method 
and the approximate Bayesian method based on the approximate likelihood.
Here is a brief description.

From the theory in Sections \ref{ss:maxstab},
if $\bY_1,\ldots,\bY_n$ are i.i.d. copies of $\bY$ on $\real^d_+$ with a distribution in the domain of attraction of a MEVD, then
the distribution of the sequence $\{R_i/n,\bW_i, i=1,\ldots,n\}$, where $R_i=Y_{i,1}+\cdots +Y_{i,d}$
and $\bW_i=\bY_i/R_i$, converges as $n\rightarrow\infty$ to the distribution of a PPP with 
density $\der\psi(r,\bw)=r^{-2} \der r \times \der H(\bw)$.

Assume that $\bx_1, \ldots, \bx_n$ are i.i.d. observations from a random vector with an unknown distribution.
Since the aim is estimating the extremal dependence, we 
transform the data into the sample $\by_1,\ldots,\by_n$ with unit Fr\'{e}chet
marginal distributions. This is done by applying the probability integral transform, 
after fitting the marginal distributions.
Next, the coordinates of the data-points are changed from Euclidean into pseudo-polar by
the transformation
$$
r_i=y_{i,1}+\cdots+y_{i,d}\quad \bw_i=\by_i/r_i, \quad i=1,\ldots,n.
$$
Then, the sequence $\{(r_i,\bw_i),i=1,\ldots,n : r_i>r_0\}$, where $r_0>0$ is a large threshold,
comes approximately from a Poisson point process with intensity
measure $\psi$. Let $\cW_{r_0}=\{(r,\bw):r>r_0\}$ be the set of points with a radial component 
larger than $r_0$, then the number of points falling in $\cW_{r_0}$ is given by 
$N(\cW_{r_0})\sim \text{Pois}\{1/\psi(\cW_{r_0})\}$.
Conditionally to $N(\cW_{r_0})=m$, the
points $\{(r_{(i)},\bw_{(i)}),i=1,\ldots,m\}$ are i.i.d. with common density 
$\der\psi(r,\bw)/\psi(\cW_{r_0})$. If we assume that $H$ is known apart
from a vector of unknown parameters $\btheta\in \Theta\subset\real^p$, 
then the approximate likelihood of the excess is
\begin{align}\label{eq:aplik}
L( \btheta; (r_{(i)}, \bw_{(i)}),i=1,\ldots,m) 
&= \frac{e^{-\psi (\cW_{r_0})} \psi(\cW_{r_0})^m}{m!}
\prod_{i=1}^m \frac{\der \psi(r_{(i)}, \bw_{(i)})}{\psi(\cW_{r_0})}\nonumber\\
&\propto \prod_{i=1}^m h( \bw_{(i)}, \btheta ),
\end{align}
where $h$ is a parametric angular density function \citep[e.g.][pp. 170--171]{engelke2012, beirlant2006}. 
In the next section the angular density models described in Section \ref{s:par_models} 
are fitted to the data by the maximization of the likelihood
\eqref{eq:aplik}. 
\index{Approximate likelihood}
For brevity the asymmetric logistic model is not considered since it has too many parameters.
The likelihood \eqref{eq:aplik} is proportional to the product of angular densities, therefore 
the maximizer of \eqref{eq:aplik} is obtained equivalently
by maximizing the log-likelihood
\begin{equation}\label{eq:aploglik}
\ell(\btheta)=\sum_{i=1}^{m}\log h(\bw_{(i)},\btheta).
\end{equation}
Denote by $\widehat{\btheta}$ the maximizer of $\ell$ and by
$\ell'(\btheta) = \nabla_{\btheta}\,\ell(\btheta)$ the score function. Since
\eqref{eq:aplik} provides an approximation of the true likelihood, then 
from the theory on model misspecification \citep[e.g.][pp. 147--148]{davison03}
it follows that 
\begin{equation*}\label{eq:godambe}
\sqrt{n}(\widehat{\btheta}-\btheta)\stackrel{d}{\rightarrow} \mathcal{N}_p(\bzero,J(\btheta)^{-1}\,K(\btheta)\,J(\btheta)^{-1}),
\quad n\rightarrow\infty,
\end{equation*}
where $\mathcal{N}_p(\bmu,\Sigma)$ is the $p$-dimensional normal distribution with
mean $\bmu$ and covariance $\Sigma$, $\btheta$ is 
the true parameter and
$$
J(\btheta)=-\E\lbrace\nabla_{\btheta} \ell'(\btheta)\rbrace,\qquad
K(\btheta)=\var_{\btheta} \lbrace \ell' ( \btheta ) \rbrace, 
$$
are the sensitive and variability matrices \citep{varin2011}.
In the case of misspecified models, model selection can be performed by computing the
Takeuchi Information Criterion (TIC) \citep[e.g.][]{akaike}, that is
\begin{align*}
TIC = -2 \left[\ell( \widehat{\btheta} ) 
- \tr \lbrace  K ( \widehat{\btheta} ) 
J^{-1} ( \widehat{\btheta} ) \rbrace\right], 
\end{align*}
where the log-likelihood, the variability and sensitive matrices are evaluated at $\widehat{\btheta}$.
The model with the smallest value of the $TIC$ is preferred.

In order to derive an approximate posterior distribution 
for the parameters of an angular density, the approximate likelihood \eqref{eq:aplik} can be
used within the Bayesian paradigm \citep[see][]{sabourin2013}. 
\index{Approximate Bayesian}
Briefly, let $q(\btheta)$ be a prior distribution on
$\btheta$, then the posterior distribution of the angular density's parameters is
\begin{equation}\label{eq:posterior}
q(\btheta|\bw)=\frac{\prod_{i=1}^m h( \bw_{(i)}, \btheta )\,q(\btheta)}
{\int_{\Theta} \prod_{i=1}^m h( \bw_{(i)}, \btheta )\,q(\btheta)\, \der\,\btheta}.
\end{equation}
With the angular density models in Section \eqref{s:par_models} the analytical expression of $q(\btheta|\bw)$ can not be derived. Therefore, we use a Markov Chain
Monte Carlo method for sampling from an approximation of $q(\btheta|\bw)$.
Specifically, we use a Metropolis--Hastings simulating algorithm \citep[e.g.][]{hastings1970}.
With the pairwise beta models we use the prior distributions described by
\citet{sabourin2013}. With the titled Dirichlet and H\"{u}sler-Reiss model we use independent zero-mean normal prior distributions 
with standard deviations equal to $3$ for $\log \alpha_j$ and $\log\,\lambda_{i,j}$ with $i,j\in I$. 
For the extremal-$t$ model we use independent zero-mean normal prior distributions 
with standard deviations equal to 3 for $\text{sign}(\rho_{ij})\text{logit}(\rho_{ij}^2)$ with $i,j\in I$, where
$\text{sign}(x)$ is the sign of $x$ for $x\in \real$ and $\text{logit}(x)=\log(x/(1-x))$ for $0\leq x\leq1$, 
and a zero-mean normal prior distribution with standard deviations equal to 3 for $\log \nu$.
Similar to \citet{sabourin2013}, for each models' parameter we select
a sample of $50\times 10^3$ observations from the approximate posterior,
after a burn-in period of length 
$30\times10^3$. These sizes have been determined using 
the Geweke convergence diagnostics \citep{geweke1992} and
the Heidelberger and Welch test \citep{heidelberger1981} respectively.

Model selection is performed using the Bayesian Information Criterion (BIC)
\citep[e.g.][]{akaike}, that is
\begin{align*}
BIC = -2\, \ell(\widehat{\btheta}) + p\{\log m+ \log(2\pi)\},
\end{align*}
where $p$ is the number of parameters 
and $m$ is the sample size. The model with the smallest value of  the BIC is preferred.


%
\section{Real data analysis: Air quality data}\label{sec:application}

We analyze the extremal dependence of the air quality data, recorded in the city centre of Leeds, UK. 
The aim is to estimate the probability that multiple pollutants will be simultaneously high in the near future. 
This dataset has been previously studied by \citet{heffernan2004}, \citet{boldi2007} and \citet{cooley2010}.
The data are the daily maximum of five air pollutants: particulate matter (PM10), nitrogen oxide (NO), nitrogen dioxide (NO2), ozone (03), and sulfur dioxide (SO2). 
Levels of the gases are measured in parts per billion, and those of PM10 in micrograms per cubic meter.
We focus our analysis on the winter season (from November to February) from 
1994 to 1998.

A preliminary analysis focuses on the data of triplets of variables. For brevity we only report the results
of the most dependent triplets: PM10, NO, SO2 (PNS), NO2, SO2, NO (NSN) and PM10, NO, NO2
(PNN). 
For each variable, the empirical distribution function is estimated with the data below the 0.7 quantile 
and a GPD is fitted to the data above the quantile \citep{cooley2010}.
Then, each marginal distribution is transformed into a unit Fr\'{e}chet. 
The coordinates of the data-points are transformed to radial distances 
and angular components. For each triplet, the 100 observations with the largest radial distances 
are retained.
\begin{table}[h!]
\setlength{\tabcolsep}{4pt}
\begin{center}
\begin{tabular}{ @{} l @{} c @{} c c c c @{}c @{}c@{} }
Model $$ & Method &\multicolumn{4}{l}{Estimates} & $\ell(\widehat{\btheta})$ & TIC/BIC \\
\hline
 TD & &$\widehat{\alpha}_{1}$ & $\widehat{\alpha}_{2}$ & $\widehat{\alpha}_{3}$\\
\hline
 PNS & L &$1.20 (0.24)$ & $0.67 (0.07)$ & $0.41 (0.08)$ &  & $199.63$ & $-399.21$ \\
& B&$1.22 (0.25)$ & $0.68 (0.11)$ & $0.42 (0.09)$ &  & & $-379.90$ \\
 NSN & L& $0.85 (0.12)$ & $0.39 (0.08)$ & $0.90 (0.11)$ &  & $200.84$ & $-401.63$ \\
 & B&$0.86 (0.15)$ & $0.39 (0.09)$ & $0.81 (0.15)$ &  &  & $-382.32$ \\
 PNN &L &$1.43 (0.28)$ & $1.55 (0.31)$ & $1.28 (0.20)$ &  & $186.35$ & $-372.64$ \\
 & B&$1.45 (0.30)$ & $1.57 (0.28)$ & $1.29 (0.23)$ &  &  & $-353.36$\\ 
 \hline
PB && $\widehat{\beta}_{1,2}$ & $\widehat{\beta}_{1,3}$ & $\widehat{\beta}_{2,3}$ & $\widehat{\alpha}$ \\
\hline
 PNS & L& $3.21 (0.70)$ & $0.47 (0.05)$ & $0.45 (0.04)$ & $0.68 (0.06)$ & $95.95$ & $-191.87$ \\
 & B&$3.31 (1.13)$ & $0.48 (0.11)$ & $0.46 (0.10)$ & $0.68 (0.09)$ & & $-166.10$ \\
 NSN & L& $0.40 (0.03)$ & $3.74 (1.77)$ & $0.50 (0.05)$ & $0.64 (0.05)$ & $102.59$ & $-205.13$ \\
 & B&$0.40 (0.09)$ & $4.00 (1.72)$ & $0.51 (0.12)$ & $0.64 (0.08)$ & &$-179.36$ \\
 PNN & L& $3.75 (1.38)$ & $0.71 (0.09)$ & $3.18 (1.21)$ & $1.35 (0.18)$ & $84.31$ & $-168.55$ \\
 & B&$3.83 (1.75)$ & $0.72 (0.16)$ & $3.70 (1.80)$ & $1.37 (0.20)$  &  & $-142.66$ \\
 \hline
 HR && $\widehat{\lambda}_{1,2}$ & $\widehat{\lambda}_{1,3}$ & $\widehat{\lambda}_{2,3}$ \\
 \hline
 PNS & L& $0.65 (0.06)$ & $0.90 (0.04)$ & $0.98 (0.03)$ & & $234.51$ & $-468.93$ \\
 & B&$0.65 (0.04)$ & $0.90 (0.04)$ & $0.98 (0.04)$ & & & $-449.67$ \\
 NSN & L& $1.00 (0.04)$ & $0.56 (0.04)$ & $0.96 (0.04)$ & & $251.80$ & $-503.54$ \\
 & B&$1.00 (0.04)$ & $0.57 (0.03)$ & $0.97 (0.04)$ & & & $-484.25$ \\
  PNN & L& $0.60 (0.05)$ & $0.70 (0.04)$ & $0.51 (0.03)$ &  & $198.23$ & $-396.38$ \\
 & B&$0.60 (0.03)$ & $0.70 (0.04)$ & $0.51 (0.03)$ &  &  & $-377.11$\\ 
 \hline
 ET && $\widehat{\rho}_{1,2}$ & $\widehat{\rho}_{1,3}$ & $\widehat{\rho}_{2,3}$ & $\widehat{\nu}$ \\ 
 \hline
 PNS & L& $0.87 (0.02)$ & $0.74 (0.03)$ & $0.66 (0.03)$ & $3.89 (0.51)$ & $152.13$ & $-304.18$ \\
 & B&$0.87 (0.02)$ & $0.77 (0.02)$ & $0.72 (0.01)$ & $4.02 (0.35)$ & &$-275.13$ \\
 NSN & L& $0.58 (0.04)$ & $0.87 (0.02)$ & $0.64 (0.03)$ & $3.50 (0.01)$ & $141.92$ & $-283.80$ \\
 & B&$0.72 (0.01)$ & $0.89 (0.02)$ & $0.73 (0.02)$ & $4.00 (0.33)$ & &$-242.50$ \\
 PNN & L& $0.88 (0.02)$ & $0.82 (0.02)$ & $0.89 (0.01)$ & $3.70 (0.78)$ & $180.74$ & $-361.38$ \\
 & B&$0.86 (0.02)$ & $0.78 (0.03)$ & $0.87 (0.02)$ & $3.21 (0.43)$  &  & $-330.33$ \\
\end{tabular}
\end{center}
\caption
[Fits of extremal dependence models to the air pollution data]
{Summary of the extremal dependence models fitted to the UK air pollution data.
For each angular density model the estimation results of the triplets of pollutants are reported.  
L and B denote the approximate likelihood and Bayesian inferential method.
Estimates are maximum likelihood (standard errors) and posterior means (standard deviations). 
}
\label{table:result_ALLdata}
\end{table}
The angular density models in Section \ref{s:par_models} are fitted to the data using the 
methods in Section \ref{sec:estimation}. 

The results are presented in Table \ref{table:result_ALLdata}. 
\begin{figure}[h!]
\begin{center}$
\begin{array}{c}
\includegraphics[width=0.8\textwidth]{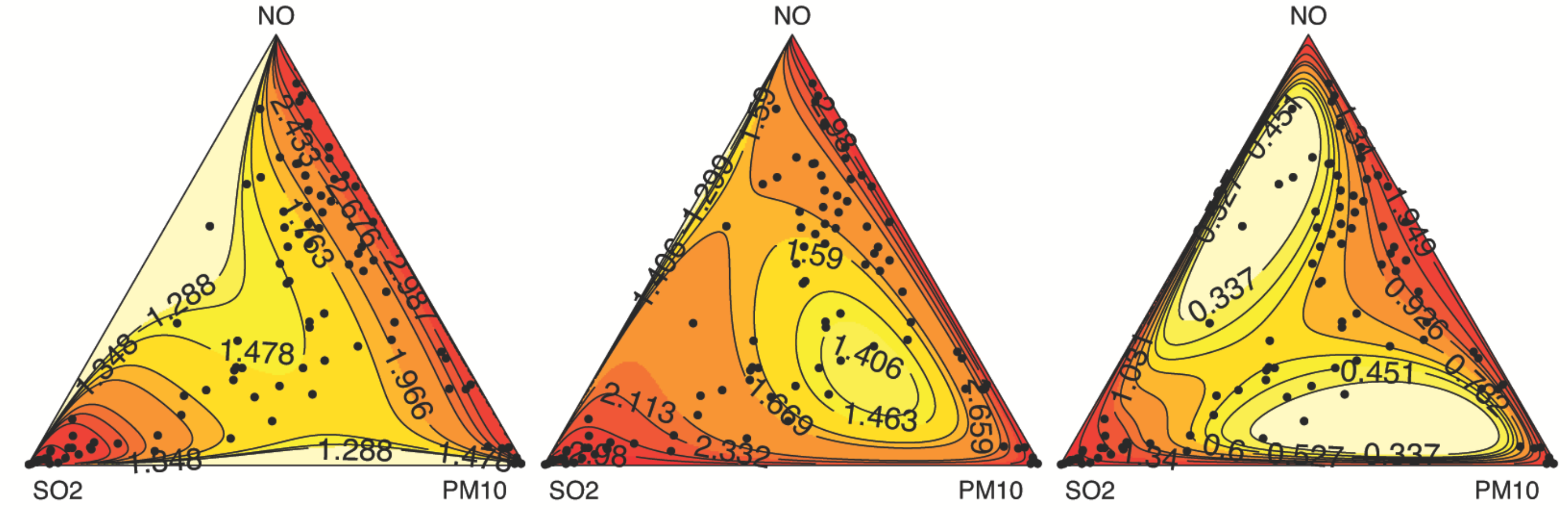} \\
\includegraphics[width=0.8\textwidth]{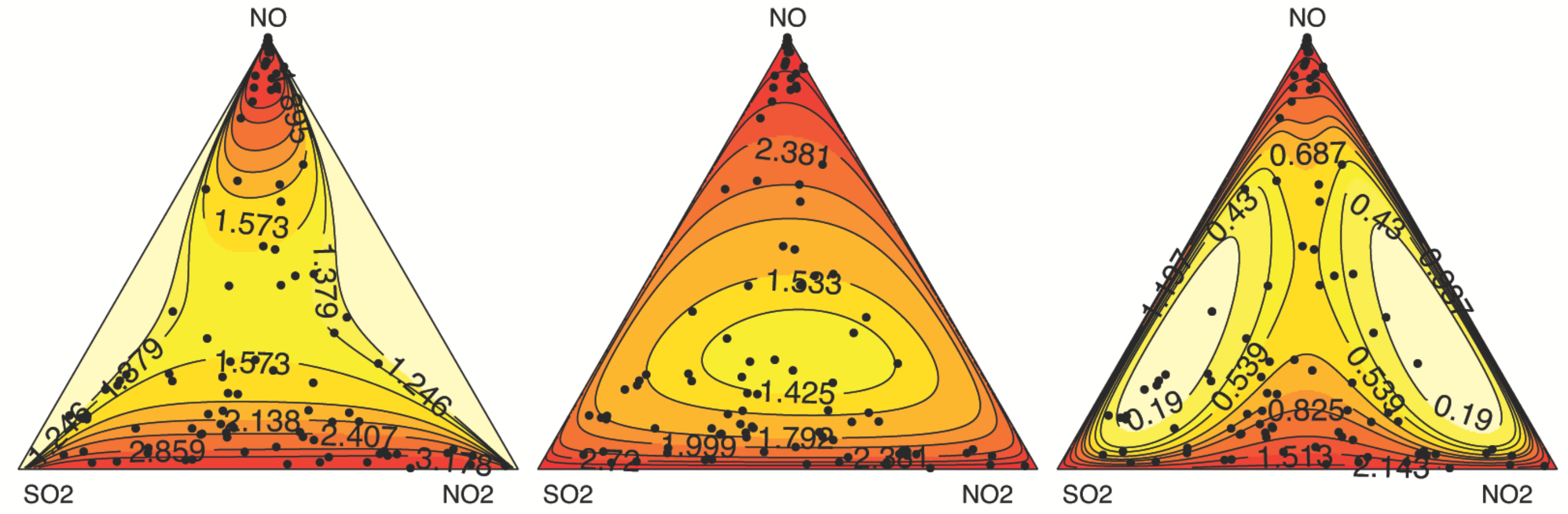} \\
\includegraphics[width=0.8\textwidth]{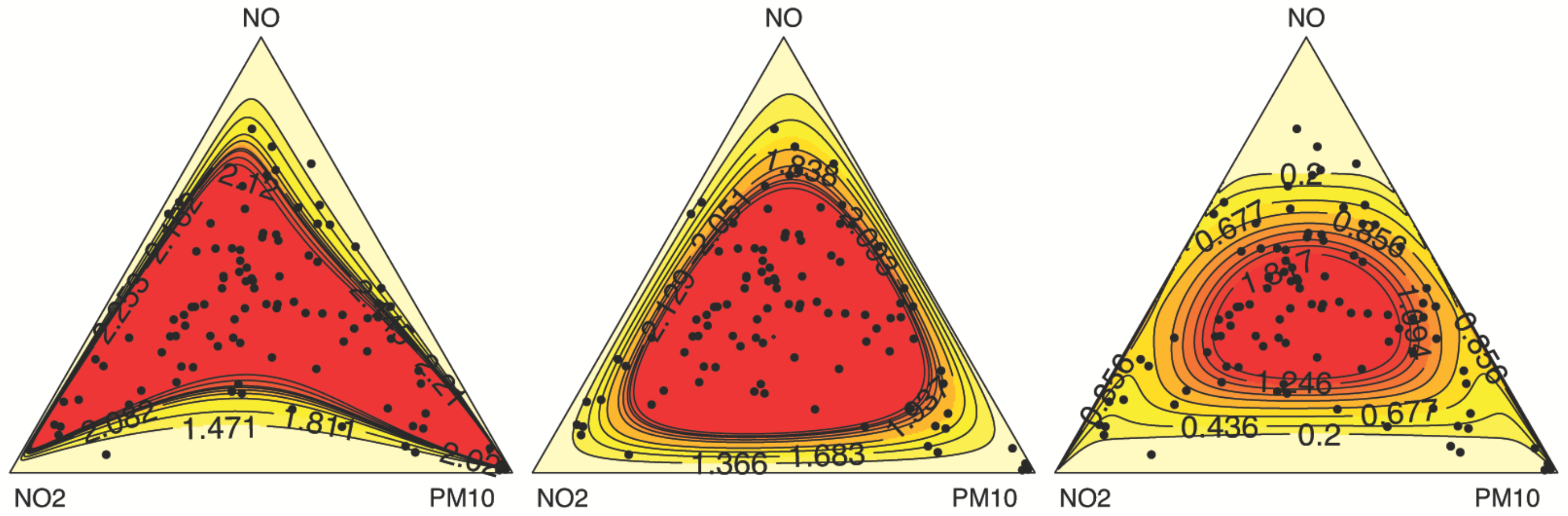}
\end{array}$
\end{center}
\caption
[Estimated angular densities]
{Estimated angular densities in logarithm scale. Dots represent the largest 100 observations.}
\label{fig:summary_fitted_densities}
\end{figure}
Maximum likelihood estimates are similar to the estimated posterior means and the estimated posterior standard deviations are typically larger than the standard errors.
For PNS we obtain the same maximum likelihood estimates
as \citet{cooley2010} with the pairwise beta model, 
however we use \eqref{eq:godambe} to compute the variances of the estimates 
and so we attain larger standard errors than they do. 
Both the TIC and BIC lead to the same model selection. The 
H\"{u}sler-Reiss model provides the best fit for all the groups of pollutants. 

From top to bottom, Figure \ref{fig:summary_fitted_densities} displays the angular densities, 
computed with the posterior means. 
From left to right the H\"{u}sler-Reiss, the tilted Dirichlet and the pairwise beta densities are reported.
With PNS, we see that there are many observations along the edge that link PM10 and NO, revealing
strong dependence between these two pollutants. There are also several observations on the SO2 vertex, reflecting 
that this pollutant is mildly dependent with the other two. 
There are also some data in the middle of the simplex, indicating that
there is mild dependence among the pollutants. 
Similarly, with NSN we see that there is strong
dependence between NO and NO2, because there are many observations along the edge that link them.
There is a mild dependence between SO2 and the other pollutants, because there is a considerable amount 
of data on the O3 vertex. Overall, there is mild dependence among the pollutants, because 
there is a small amount of data in the middle of the simplex. 
With PNN we see that most of the observations are placed on the middle of the simplex revealing 
an overall strong dependence among the pollutants. There is a small amount of data along the edge that link
NO2 and NO and on the PM10 vertex. This reflects more dependence between NO2 and NO than between
NO2 and PM10  and PM10 and NO.
All these features are well captured by the angular densities estimated using the H\"{u}sler-Reiss model.

%
%
%
 With this analysis we found that O3 is only weakly dependent with the other pollutants. This result was
also found by \citet{heffernan2004}. Then, the second part of the analysis focuses only on PM10, NO, NO2 and SO2.
Now, because a larger number of parameters needs to be estimated,
then the 200 observations with the largest radial distances are selected \citep[see][]{cooley2010}. 
Table \ref{table:result_FULL} presents the estimation results.  
For brevity we only report the maximum value of the log-likelihood,  
the TIC and the BIC.
\begin{table}[t!]
\begin{center}
\begin{tabular}{ c r r r r }
& Tilted Dirichlet & Pairwise Beta & H\"{u}sler-Reiss & extremal-$t$ \\ 
\hline
$\ell(\widehat{\btheta})$ & $654.3$ & $402.5$ & $762.7$ & $532.3$ \\
TIC & $-1308.6$ & $-805.0$ & $-1525.3$ & $-1064.5$ \\
BIC & $-1280.0$ & $-753.4$ & $-1475.5$ & $-974.7$ \\
\end{tabular}
\end{center}
\caption
[Fits of the full extremal dependence model to the data]
{Summary of the extremal dependence models fitted to the UK air pollution data.}
\label{table:result_FULL}
\end{table}
The H\"{u}sler-Reiss model provides the smallest values of the TIC and BIC, revealing again that it
better fits the pollution data. Accordingly hereafter calculations will be made using this model and the
estimates obtained with the Bayesian approach.

We summarize the extremal dependence of the four variables using the extremal coefficient \eqref{eq:extremalcoeff}
and the coefficient of tail dependence \eqref{eq:taildep_coef}.
Specifically, $\widehat{\vartheta}=2.267$ with
a $95\%$ credible interval is equal to $(1.942,2.602)$ and
$\widehat{\chi}= 0.242 $ with a $95\%$ credible interval is $(0.150,0.361)$. 
These results suggest a strong extremal dependence among the pollutants.
The estimated extremal dependence can be used in turn to estimate the probability that multiple pollutants exceed a high threshold.
Consider a value $\by$ whose radial component is a high threshold $r_0$. 
Then, the probability of falling in the failure region \eqref{eq:failure}
is approximately equal to the right hand side of \eqref{eq:prob_failure}.
Because the exponent function is related to the tail function
by the inclusion-exclusion principle, then using \eqref{eq:exponent_husler} we have
\begin{equation}\label{eq:est_prob_ruin_set}
\P\{Y_1>y_1,\ldots,Y_d>y_d\}\approx  \sum_{j=1}^d  \frac{1}{y_j} \bar{\Phi}_{d-1} 
\Big\lbrace \Big (\lambda_{k,j} +\frac{\log y_k/y_j }{2\lambda_{k,j}} \Big )_{k \in I_j}; \bar{\Lambda}_j \Big\rbrace,
\end{equation}
where $\bar{\Phi}_{d-1}$ is the survival function of the multivariate normal distribution \citep{nikoloulopoulos2009}.
Similar to \citet{cooley2010} we define three extreme events: $\{\text{PM}10 > 95, \text{NO} > 270, \text{SO}2 > 95\}$, 
$\{\text{NO}2 > 110, \text{SO}2 > 95, \text{NO} > 270\}$  and
$\{\text{PM}10 > 95, \text{NO} > 270, \text{NO}2 > 110, \text{SO}2 > 95\}$. Then, we
compute probability \eqref{eq:est_prob_ruin_set} using in place of the parameters their estimates.
Table \ref{table:proba_excess} reports the results. 
For the three events the estimates fall inside the $95\%$ confidence intervals highlighting the 
ability of the model to estimate such extreme events.
\begin{table}[t!]
\begin{center}
\begin{tabular}{ l l l l }
 & Event $1$ & Event $2$ & Event $3$ \\
\hline
 Excess / $n$  & $18 /528 $ & $14 / 562$ & $12 /528$ \\
 Emp. Est. & $0.034\; ( 0.019, 0.050 )$
  & $0.025\; ( 0.012, 0.038 )$ & $0.023\; ( 0.010 , 0.035 )$ \\
 Mod. Est. & $0.038$ & $0.030$ & $0.030$ \\ 
\end{tabular}
\end{center}
\caption
[Probability estimates of excesses]
{Probability estimates of excesses. The first row reports the number of excess and the sample size.
The second row reports the empirical estimates and between brackets 
the $95\%$ confidence intervals obtained 
with the normal approximation. The third row reports the model estimates.}
\label{table:proba_excess}
\end{table}
\begin{figure}[b!]
\begin{center}$
\begin{array}{ccc} 
\includegraphics[width=0.3 \textwidth]{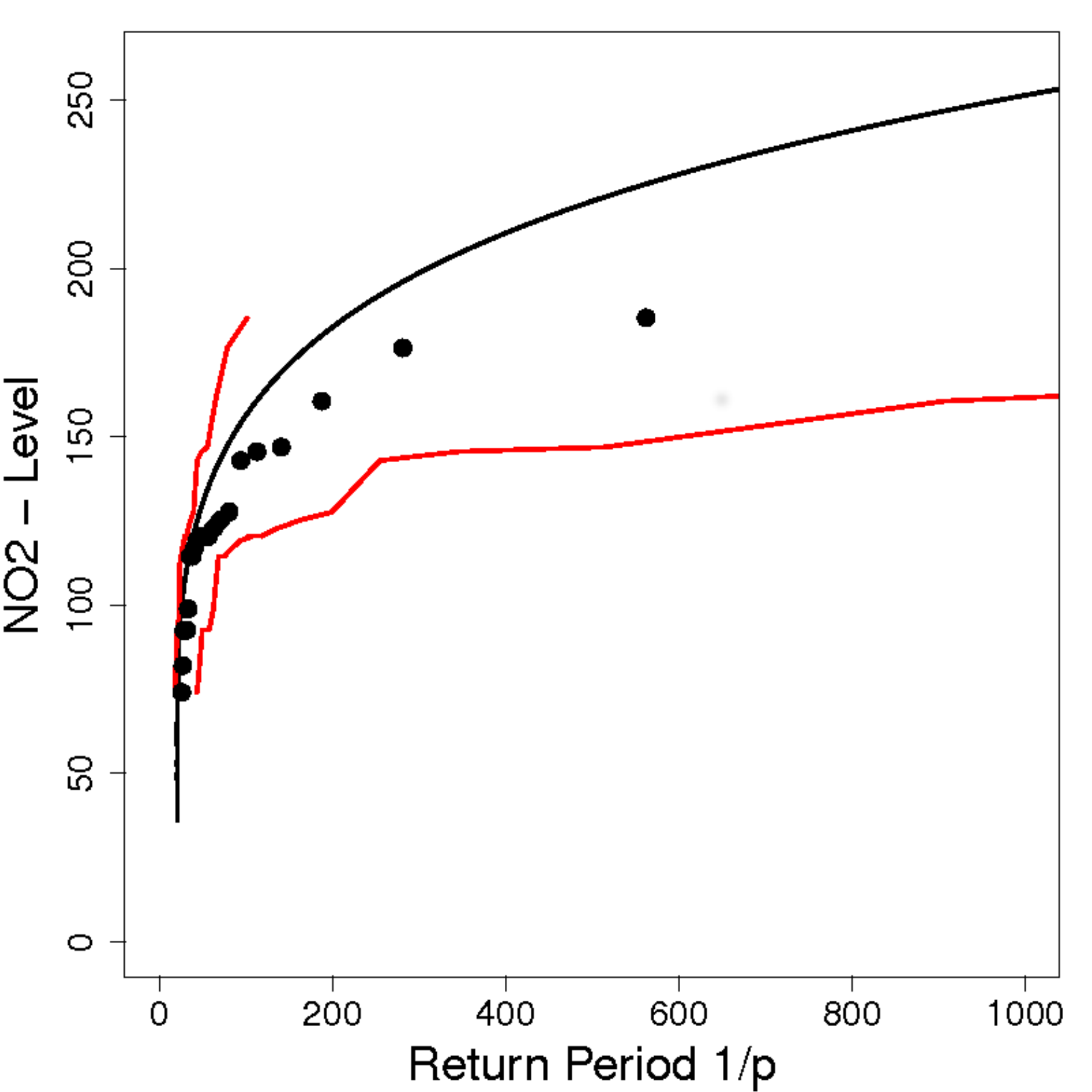} 
\includegraphics[width=0.3 \textwidth]{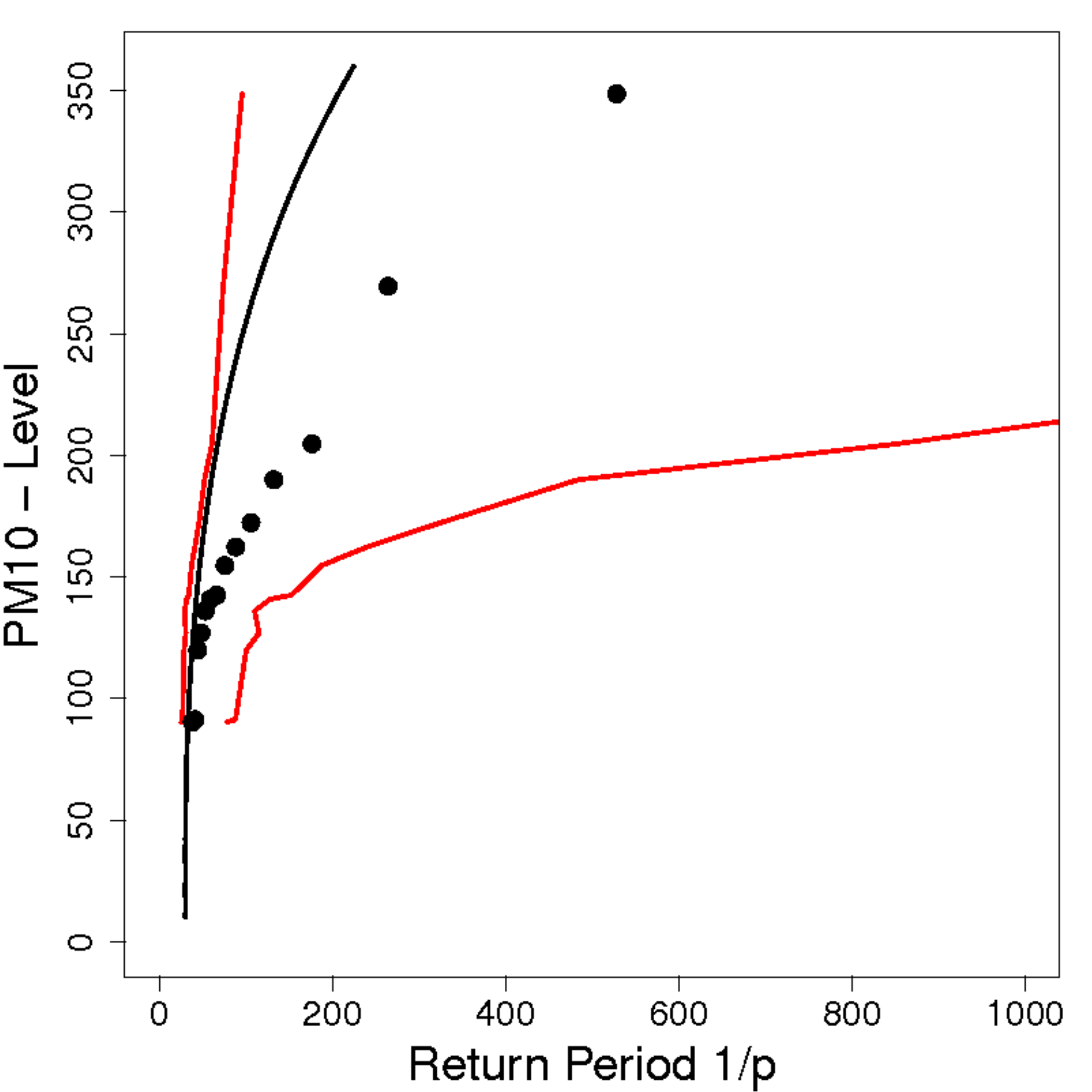}\\
\includegraphics[width=0.3 \textwidth]{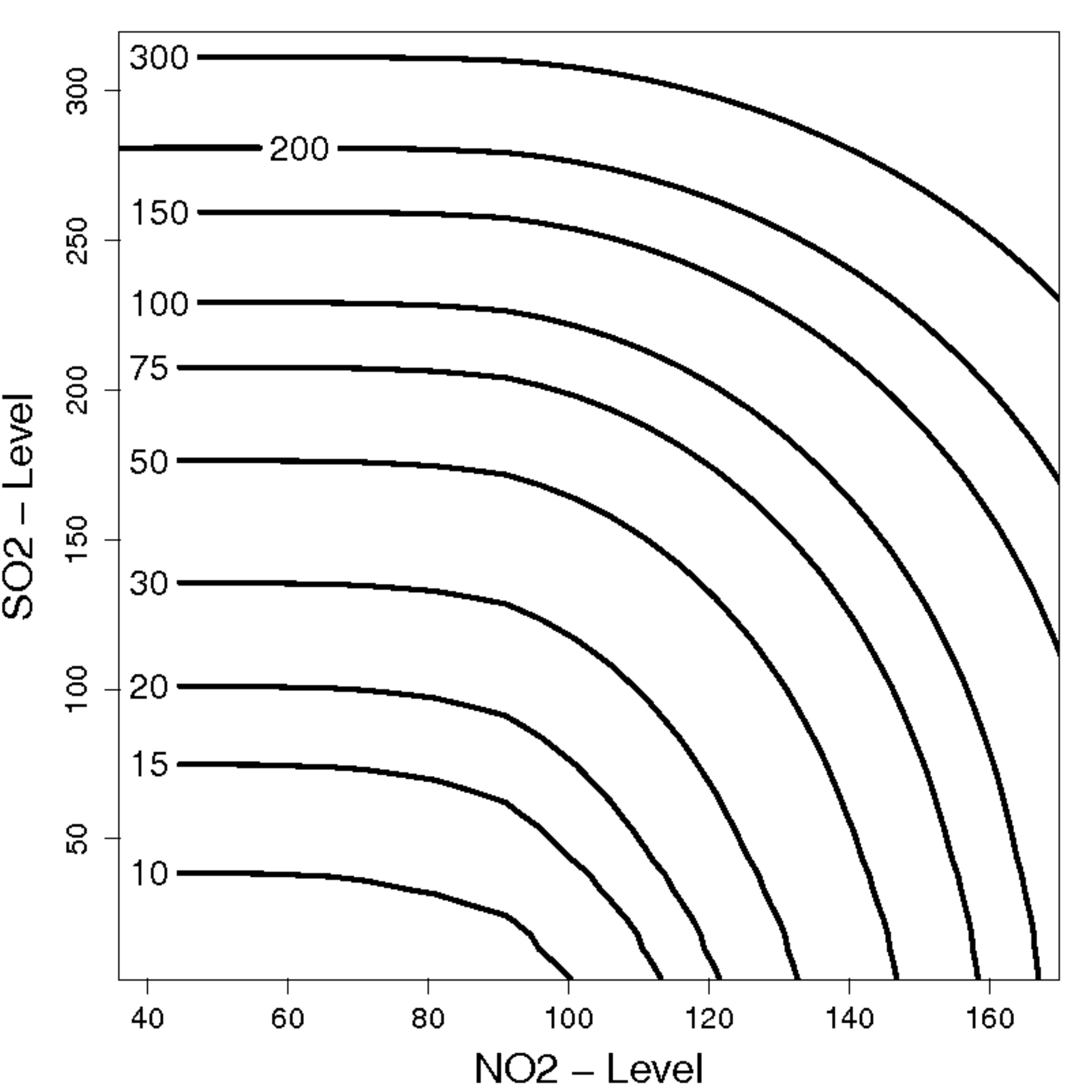} 
\includegraphics[width=0.3 \textwidth]{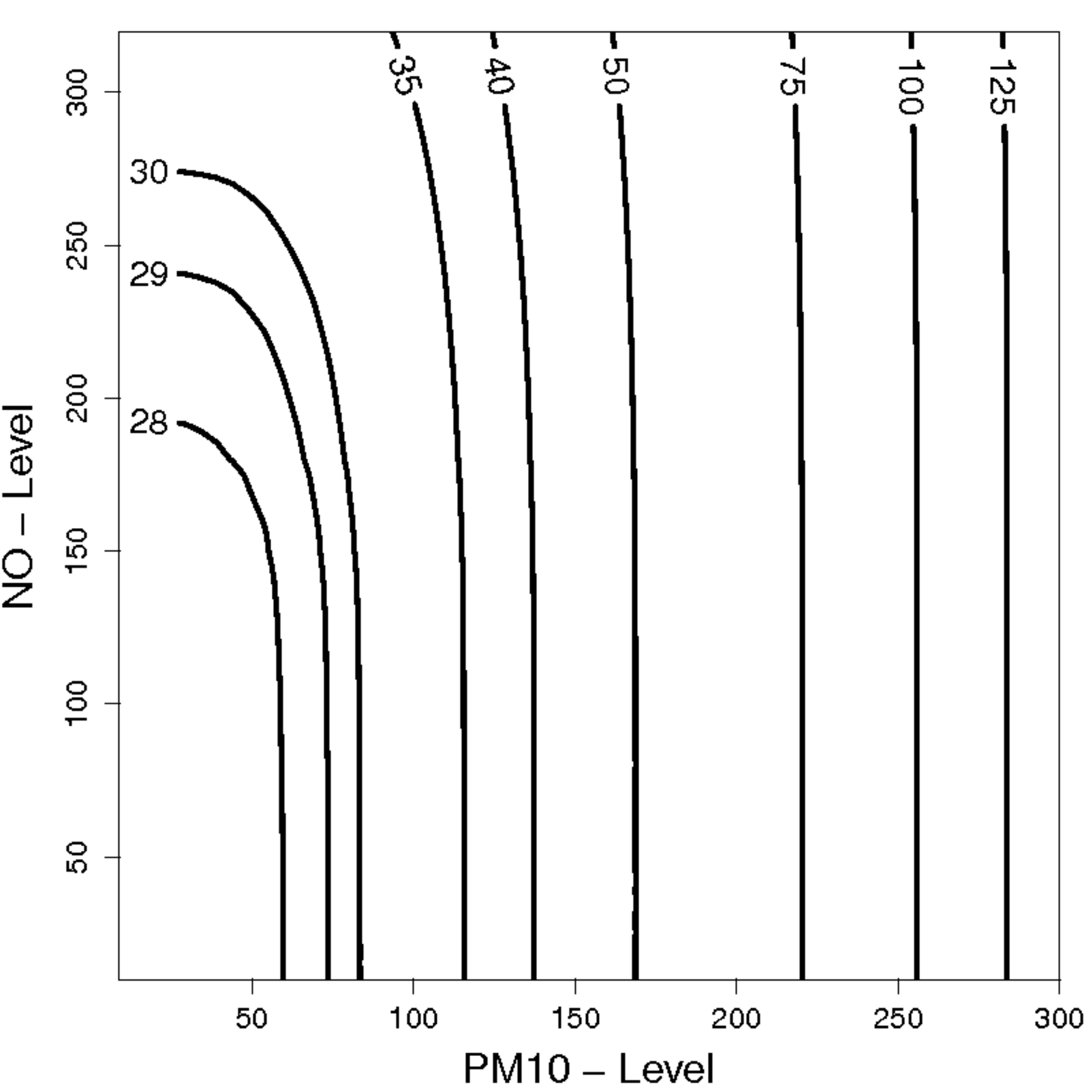}
\end{array}$
\end{center}
\caption
[Joint return level plots]
{Joint return level plots of single components NO2 and PM10 and of the two components
(SO2, NO2) and (NO, PM10)}
\label{fig:return_levels}
\end{figure}

The right-hand side of \eqref{eq:est_prob_ruin_set} can also be used for estimating 
joint return levels. In the univariate case see \citet[][pp.49--50]{coles2001}. 
In the multivariate case different definitions of return levels 
may be available \citep{Johansen04}.
Let $J\subset I$, $\{x_i, i\in I\backslash J\}$ be a sequence of fixed high thresholds
and $p\in(0,1)$ be a fixed probability.
Given a return period $1/p$, we define {\it joint return levels}
the quantiles $\{y_{j;p}, j\in J\}$ that satisfy the equation
$$
p=\P(Y_j>y_{j;p}, Y_i>x_i, \, j\in J,i\in I\backslash J). 
$$
Figure \ref{fig:return_levels} displays univariate and bivariate {\it joint return level plots}. 
When $J=\{j\}$, with $j\in I$, then the joint return level plot displays $y_{j;p}$ against $1/p$
for different values of $p$. When $J=\{i,j\}$, with $i,j\in I$, then for different values of $1/p$ 
the contour levels of $(y_{i;p},y_{j;p})$ are displayed.
With solid lines, the top-left and right panels of Figure \ref{fig:return_levels} report the
estimated return levels of NO2 and PM10 jointly to the extreme events $\{\text{SO}2 > 95, \text{NO} > 270\}$ 
and $\{\text{NO} > 270, \text{NO}2 > 110, \text{SO}2 > 95\}$ respectively.  
The dots are the empirical estimates and the red solid lines are the
pointwise $95\%$ confidence intervals. These are computed using the normal approximation when $p>0.02$ 
and using exact binomial confidence intervals when $p<0.02$. 
The bottom-left and right panels report the contour levels of the
return levels for (NO2, SO2) and (PM10, NO) jointly to events $\{\text{NO} > 270\}$ 
and $\{\text{NO}2 > 110, \text{SO}2 > 95\}$ respectivelly.

The joint return level can be interpreted as follows. For example, from the top-right panel 
we have that the 50 years joint return level of PM10 is 166. Concluding, 
we expect that PM10 will exceed the level 166 together with the event that
NO, SO2 and NO2 simultaneously exceed the levels 270, 95 and 110 respectively, on average
every 50 years.
\section{Computational details}\label{comp}
The figures and the estimation results have been obtained using the free software {\tt R}  \citep{rteam10} and
in particular the package {\tt ExtremalDep}, available at 
{\tt https://r-forge.r-project.org/projects/extremaldep/}. Bayesian estimation is obtained
using and extending some routines of the package {\tt BMAmev}.
The left and middle panels of Figure \ref{fig:extremal_dep} were obtain using the 
routines {\tt scatter3d} and {\tt polygon3d} of the package {\tt plot3D}.

\bibliographystyle{asa}
\bibliography{out_ext_dep.bib}

\end{document}